\documentclass[journal]{IEEEtran}
\usepackage{graphicx}
\usepackage{cite}
\usepackage{picinpar}
\usepackage{amsmath}
\usepackage{url}
\usepackage{flushend}
\usepackage[latin1]{inputenc}
\usepackage{colortbl}
\usepackage{soul}
\usepackage{multirow}
\usepackage{pifont}
\usepackage{color}
\usepackage{alltt}
\usepackage[hidelinks]{hyperref}
\usepackage{enumerate}
\usepackage{siunitx}
\usepackage{breakurl}
\usepackage{epstopdf}
\usepackage{pbox}
\usepackage{braket}
\usepackage{stfloats}
\usepackage{amsmath}
\usepackage{amssymb}
\usepackage{blindtext}
\usepackage{subfig}
\usepackage{tabu}
\usepackage{multirow}
\usepackage{comment}

\usepackage{amsmath}
\usepackage{amsthm}
\usepackage{amssymb}
\usepackage{mathrsfs,bm}
\usepackage{graphicx}
\usepackage[linesnumbered,ruled,vlined]{algorithm2e}

\usepackage{booktabs}
\usepackage{multirow}
\usepackage{siunitx}
\usepackage{algorithmicx}

\usepackage{algpseudocode} %for algorithms
\usepackage{multirow}
\usepackage{bbm}
\usepackage{arydshln}
\usepackage{wrapfig}
\usepackage[normalem]{ulem}
\usepackage{tabu}
\usepackage{multicol}
\usepackage{tikz}
\usetikzlibrary{arrows.meta, positioning}
\usepackage{makecell}
\usepackage{tikz}
\newcommand{\ignore}[1]{}

\newcommand{\name}{\text{PhotonIDs}}

\usepackage{amsmath, amssymb, amsthm}
\usepackage[normalem]{ulem}

\begin{document}
% \title{Q-NET: Neural Network enabled Qubit Detection and Error Correction for Quantum Communication}
\title{PhotonIDs: ML-Powered Photon Identification System for Dark Count Elimination}

\author{
	\vskip 1em
	
	Karl C. Linne (Kai Li), Sho Uemura, Yue Ji, Allen Zang,  Ian Chin, Martin Di Federico, Gustavo Cancelo, Orlando Quaranta, Debashri Roy
 \\
 % \IEEEauthorblockN{\footnotesize *Kai Li, and Debashri Roy have equally contributed to this work.}

	\thanks{
	
% 		Manuscript received Month xx, 2xxx; revised Month xx, xxxx; accepted Month x, xxxx.
% 		This work was supported in part by the xxx Department of xxx under Grant  (sponsor and financial support acknowledgment goes here).
		
		 Karl C. Linne (Kai Li), Allen Zang, and Tian Zhong are with Pritzker School of Molecular Engineering,  University of Chicago, Chicago, 60637 , USA. (e-mail: karlchrislinne@gmail.com). 
         \\

         Sho Uemura, Martin Di Federico, and Gustavo Cancelo are with Fermi National Accelerator Laboratory, Batavia, 60510, USA.
         \\

         Yue Ji with New York University, New York, 10012; USA. Orlando Quaranto is with Argonne National Laboratory, Lemont, 60439; and Debashri Roy is with The University of Texas at Arlington.
		
% 		Third C. Author3 is with the National Institute of xxx, City, Zip code, Country (corresponding author to provide phone: xxx-xxx-xxxx; fax: xxx-xxx-xxxx; e-mail: author@ domain.gov).
	}
}

\maketitle

\begingroup
\renewcommand\thefootnote{}\footnotetext{\footnotesize \textit{This work has been submitted to the IEEE Journal on Selected Areas in Communications for possible publication. Copyright may be transferred without notice, after which this version may no longer be accessible.}}\addtocounter{footnote}{-1}
\endgroup

\begin{abstract} 
Reliable single photon detection is the foundation for practical quantum communication and networking. 
However, today's superconducting nanowire single photon detector(SNSPD) inherently fails to distinguish between genuine photon events and dark counts, leading to degraded fidelity in long-distance quantum communication. In this work, we introduce \name, a machine learning-powered photon identification system that is the first end-to-end solution for real-time discrimination between photons and dark count based on full SNSPD readout signal waveform analysis. \name ~demonstrates: 1) an FPGA-based high-speed data acquisition platform that selectively captures the full waveform of signal only while filtering out the background data in real time; 2) an efficient signal preprocessing pipeline, and a novel pseudo-position metric that is derived from the physical temporal-spatial features of each detected event; 3) a hybrid machine learning model with near 98\%  accuracy achieved on photon/dark count classification. Additionally, proposed \name ~ is evaluated on the dark count elimination performance with two real-world case studies: (1)  20 km quantum link, and (2) Erbium ion-based photon emission system. Our result demonstrates that \name ~could improve more than 31.2 times of signal-noise-ratio~(SNR) on dark count elimination. \name ~ marks a step forward in noise-resilient quantum communication infrastructure.  
\end{abstract}


% \markboth{IEEE TRANSACTIONS ON INDUSTRIAL ELECTRONICS}%
{}

\definecolor{limegreen}{rgb}{0.2, 0.8, 0.2}
\definecolor{forestgreen}{rgb}{0.13, 0.55, 0.13}
\definecolor{greenhtml}{rgb}{0.0, 0.5, 0.0}

\section{Introduction}
\label{sec:intro}
% \IEEEPARstart{Q}{uantum} networking~\cite{kimble2008quantum,wehner2018quantum} holds the promises for long distance secure quantum communication ~\cite{gisin2007quantum,li2023bip}, distributed quantum sensing~\cite{proctor2017networked,proctor2018multiparameter,zhang2021distributed}, and interconnecting future quantum computers~\cite{gottesman1999demonstrating,cacciapuoti2019quantum,caleffi2024distributed}. 
\IEEEPARstart{L}{everaging} fundamental quantum mechanical principles such as entanglement and superposition, quantum networks~\cite{kimble2008quantum,wehner2018quantum} offer unprecedented capabilities in long-distance secure communication~\cite{gisin2007quantum,li2023bip}, networked quantum metrology~\cite{proctor2017networked,zhang2021distributed}, and quantum processor interconnection~\cite{gottesman1999demonstrating,cacciapuoti2019quantum}. 
As these applications move from theoretical models to real-world implementation, the demand for ultra-sensitive, low-error, and real-time photon detection and analysis becomes a critical bottleneck at the physical layer of quantum networks.
The exceptional performance of the SNSPD ~\cite{grunenfelder2023fast} 
characteristics, including high detection efficiency($>90\%$), low timing jitter (on the order of tens of picoseconds), and sub-nanosecond reset time, make it the most efficient detector for advanced quantum experiments and practical quantum communication systems. SNSPDs\cite{linne2025squad} are widely deployed in fiber-based quantum key distribution (QKD), entanglement-based communication protocols, photonic quantum computers, and single-photon emitter readout systems.

\begin{figure}[t!]
\centering
\includegraphics[width=1.0\linewidth]{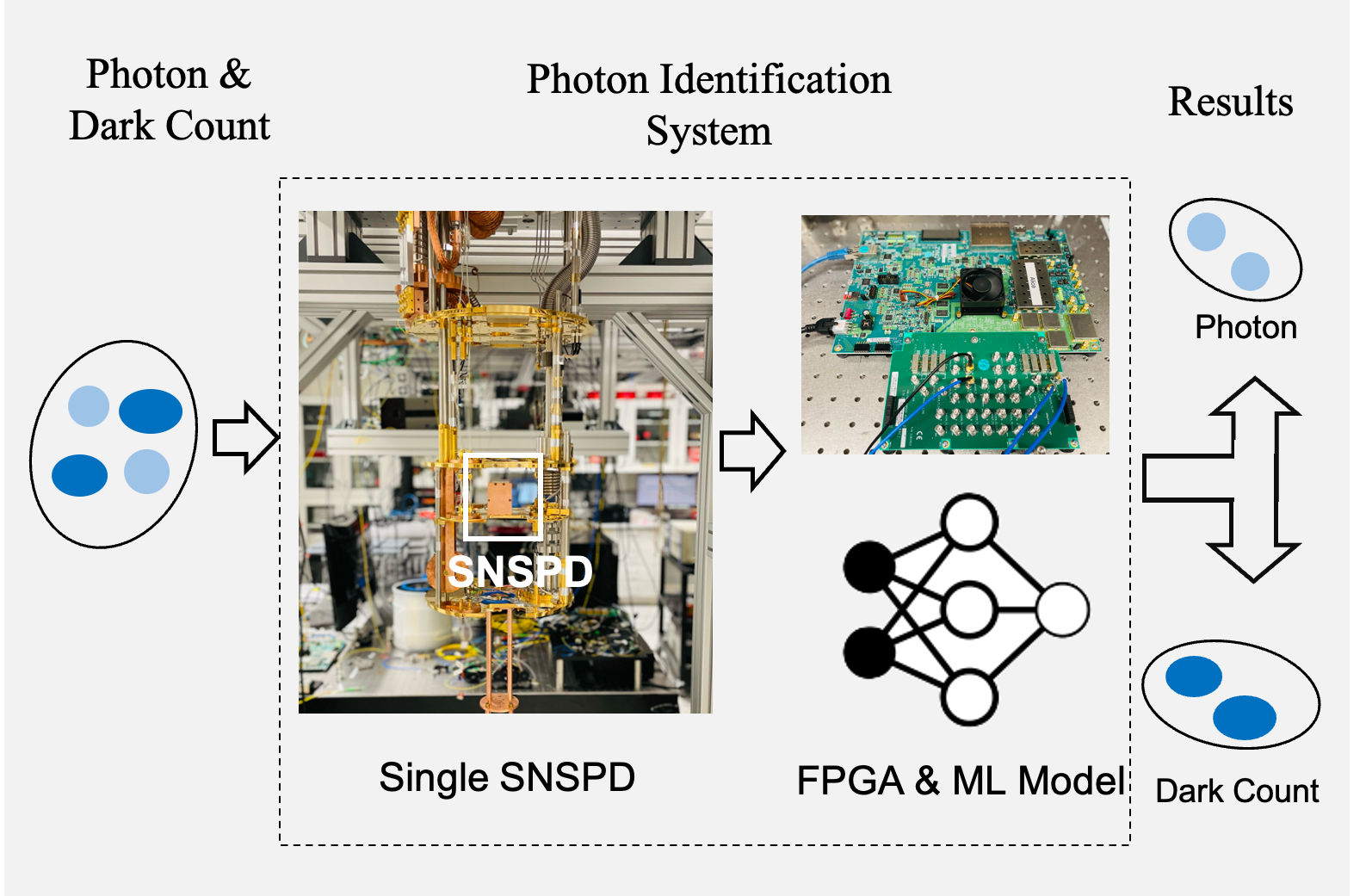}
\caption{An overview of the proposed hybrid machine learning powered photon identification system on the classification of photon and dark count. The system includes SNSPD setup, FPGA data acquisition platform, and the proposed hybrid ML model.}
\label{fig:1st}
\vspace{-6.8mm}
\end{figure}

% However, current SNSPD-based photon detection systems fundamentally operate as binary threshold detectors, providing only event-based output used to count the number of photons and arrival time. Specifically, each detection event is recorded by the readout of a RF signal pulse from the SNSPD, regardless of whether the recorded event caused by a genuine photon or by environmental noise leading to a dark count. This inability to differentiate between true photon arrivals and dark count poses increasing challenges in scenarios of photon-starved regimes or long-distance fiber transmission where environmental noise has accumulated.
% As a result, there is a growing consensus in the quantum community that \textit{dark count-free} ability to actively and accurately distinguish photons from dark counts at the detection layer is an essential requirement for quantum communication networks toward large-scale deployment.

However, today's single photon detection system has limited functionality of counting the number and recording the time of arrival of photons based on the RF signal pulse of the SNSPD readout. Additionally, almost all photon detectors operate with non-zero dark counts (false positive clicks) in the absence of input photons. Any dark count, either caused by stray photons from the environment(long fiber) or noise in the electronic circuits, would lead to erroneous detection and inaccurate readout of quantum information. This dark count will, for instance, degrade the fidelity of entanglement generation between remote quantum network nodes~\cite{lvovsky2009optical,lei2023quantum,sangouard2011quantum,azuma2023quantum} or reduce the true photon transmission rate as a system. Hence, the development of an intelligent photon identification system that simultaneously identifies true photons and performs a dark-count-free operation is highly important and will bring significant advances to quantum communications~\cite{chaudhary2023learning} and networking\cite{li2023q, linne2025q}, as well as quantum information science on a large scale.

\noindent\textbf{State-of-the-Art and Limitations:} 
To reduce dark counts and the related impacts, several approaches have been reported to suppress or eliminate spurious detection events in SNSPD-based systems. These methods span optical, structural, temporal, and operational domains, each with its own set of benefits and limitations.

In the work of Shibata et al.~\cite{luskin2024superconducting}, a narrow-band optical filter is inserted at the fiber output to suppress blackbody radiation and background noise, by which a two-order-of-magnitude reduction in dark count rate is obtained. Some work reduces the dark count rate via nanowire geometry and material optimization. For instance, the work in \cite{zhang2019saturating, baghdadi2021enhancing} utilized rounded bends and variable-thickness nanowire designs to improve the critical current and reduce the dark count rate. More statistically, the studies in \cite{chen2015dark, schapeler2024electrical} statistically distinguish photons from dark counts based on their temporal distribution, based on a huge amount of offline data analysis. Gated-mode operation method with predefined time windows synchronized with expected photon arrival times is proposed in ~\cite{akhlaghi2012gated} to reduce the probability of registering random dark counts~\cite{akhlaghi2012gated}.

The above studies have made huge progress on reducing dark counts through different methods. However, these existing methods suffer from fundamental limitations: 
\textit{1)~Fabrication and System Complexity:} The extra optical filter increases the system complexity and limits the detection bandwidth,  and the specific nanowire fabrication requires advanced fabrication processes and lacks adaptability to dynamic noise conditions.   
\textit{2)~Data loss and Time Consuming:} 
The statistical analysis needs a huge amount of data as the extra cost and offline processing increase the system's time-consuming, and the gate-mode requires precise timing control. 
\textit{3)~Binary Detection Constraints:}Today's
Solutions do not overcome the binary nature of SNSPDs, which inherently lack the capability of identifying photon event or dark count events, making them unsuitable for large-scale communication.

To address these challenges, we propose \name, a compact and intelligent system that leverages full-waveform analysis of each detected event captured from a single SNSPD. By applying a machine learning model to extract wavelength-sensitive temporal features, \name ~enables a high-accuracy photon/dark-count identification system without requiring external optical components or extensive data-driven statistical analysis. The proposed architecture provides a scalable, dark count-free solution for photon-based quantum networking across distributed nodes.

\noindent$\bullet$ \textbf{Challenges of \name:} \name ~introduces the first framework for an intelligent photon/dark count identification system using machine learning; its practical realization must overcome several hardware and signal processing challenges throughout the full waveform signal collection and processing pipeline.
\begin{itemize}
    \item \textit{(C1) Accurate Data Acquisition:} The SNSPD readout signal comes from photon or dark count events occurring on the nanosecond timescale, which demands a high-speed acquisition system capable of capturing full waveform dynamics with nanosecond timescale resolution.
    
\item \textit{(C2) Real-time Event Isolation:} \name~ must extract meaningful photon and dark count events from a continuous detector stream, rather than simply counting threshold crossings. This requires real-time suppression of baseline background data to reduce memory usage and ensure efficient data handling during high-rate acquisition.
\item \textit{(C3) Electronic Noise and Signal Integrity:}  
Readout electronics can introduce artifacts, baseline drift, and high-frequency interference into the digitized waveform, potentially obscuring the true photon signal. A robust preprocessing pipeline is required to suppress such distortions and remove the high-frequency interference.
\item \textit{(C4) Position-Dependent Signal Variability:} SNSPD readout signal waveform shape has subtle variations both in time and amplitude, caused by the different absorption between the nanowire and the photon/dark count. Accurate modeling of this spatial dependence is crucial for improving photon/dark count classification accuracy and enhancing the model's ability to generalize across diverse applications. 
   % \item \textit{(C4)Position-Dependent Signal Variability:}  Photons absorbed at different locations along the SNSPD nanowire can induce subtle but consistent variations in the resulting waveform shape. Accurately modeling this spatial dependency is essential to improve classification robustness and enhance the model’s ability to generalize across diverse detection scenarios.
\end{itemize}

\noindent$\bullet$ \textbf{\name ~Framework and Contributions:} The proposed {\name}~system is illustrated in Fig.~\ref{fig:1st}.
As shown in Fig.~\ref{fig:1st}, the proposed photon identification system consists of a single SNSPD, an FPGA-based acquisition platform, and a hybrid ML model. Each event caused by a photon or dark count is detected by the SNSPD, whose full waveforms of the readout are digitized in real time by the FPGA within a nanosecond scale. The obtained full waveform information is then analyzed by the proposed hybrid machine learning model to identify whether the recorded event is a photon or a dark count with high classification accuracy. 
By identifying photons from dark counts at the single-event level, \name enables a scalable, low-loss pathway to dark count-free quantum networking.

The main contributions of \name are summarized as following:

\begin{enumerate}
\item We develop an FPGA-based data acquisition framework to capture the complete waveform of each SNSPD readout with sub-nanosecond resolution, addressing Challenge C1.

\item To overcome C2, the proposed FPGA framework supports a real-time event trigger and event-only signal extraction approach, including counting the number of triggered events, real-time background data suppression, and spontaneous on-board data saving and streaming.

% \item We employ a chain of broadband, low-noise, high-speed RF amplifiers after the SNSPD (C3), ensuring high-fidelity signal amplification with minimal distortion.

\item To address C3, we develop a preprocessing pipeline that removes high-frequency noise and enhances the time resolution on each waveform to improve downstream ML performance.

\item To mitigate C4, we introduce a hybrid learning architecture that combines waveform morphological features with temporally informed signal patterns, enhancing the photon and dark count classification accuracy of the proposed hybrid ML model.
\end{enumerate}

The rest of the paper is organized as follows. We present the background in Sec.~\ref{sec:background} and system design in Sec.~\ref{sec:systemdesign}. We describe the theory of the proposed hybrid machine learning model in Sec.~\ref{sec:theory}. The experimental evaluation results and case studies are demonstrated in Sec.~\ref{sec:experimental_evaluation} and Sec.~\ref{sec:casestudy}. We conclude our work in last section.

\section{Background}
\label{sec:background}

This section will give a background review of the SNSPD operational mechanism and today's single photon detection technology. 

\subsection{Operational Mechanism of SNSPD}

\begin{figure}[t!]
    \centering
    \includegraphics[width=0.7\linewidth]{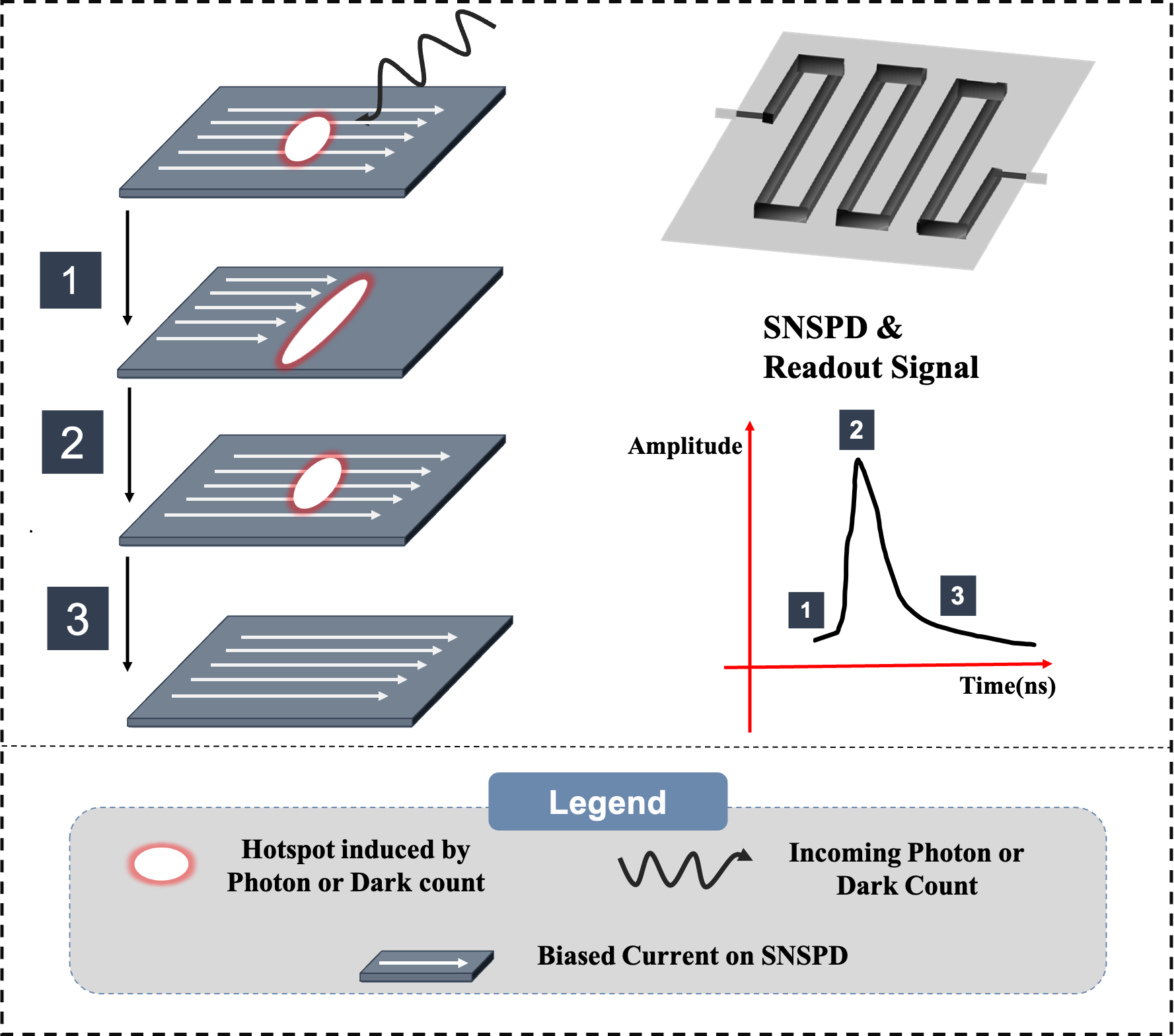}
    \label{subfig:fig:s_con}

    \caption{Operational mechanism of SNSPD and readout signal with three phases of variation.}
    \label{fig:sc_work}
    \vspace{-0.8cm}
\end{figure}

Fig.~\ref{fig:sc_work} sketches the operation mechanism of a biased meander SNSPD and the associated readout signal. When biased and operated at millikelvin temperatures, the meandered nanowire remains in the superconducting state, exhibits (nearly) zero resistance, and carries the bias current uniformly along its length. An incident photon or spontaneous dark count event perturb the nanowire and initiates an electro thermal transient that proceeds through three stages, labeled 1 to 3 in Fig.~\ref{fig:sc_work}.  In phase 1, the superconducting nanowire absorbs a photon (or dark count) and its energy. The absorbed energy creates a localized hotspot that rapidly expands across the wire width into a resistive barrier, diverting the bias current and producing the rising edge of the readout signal, which then reaches its peak. In the second phase, heat begins to dissipate into the substrate beneath the nanowire, and the readout signal decays from the peak amplitude. At the third phase, the hotspot completely disappears, allowing the nanowire to return to its superconducting state with zero resistance, making it ready to absorb the next photons. The variation of the superconducting nanowire induced by absorbed photon (or dark count) produces a measurable time-varying electrical signal with fast rising and slow decay. Specifically, the waveshape, peak value, and time duration of the readout signal are closely correlated with both the photon's properties~\cite{semenov1995analysis}\cite{natarajan2012superconducting} and the absorption position along the nanowire. This correlation is the foundation of the proposed ML-powered photon identification system in this work.

\subsection{State-of-the-Art Photon Detection systems} 
\begin{figure}[h!]
\centering
\includegraphics[width=1.0\linewidth]{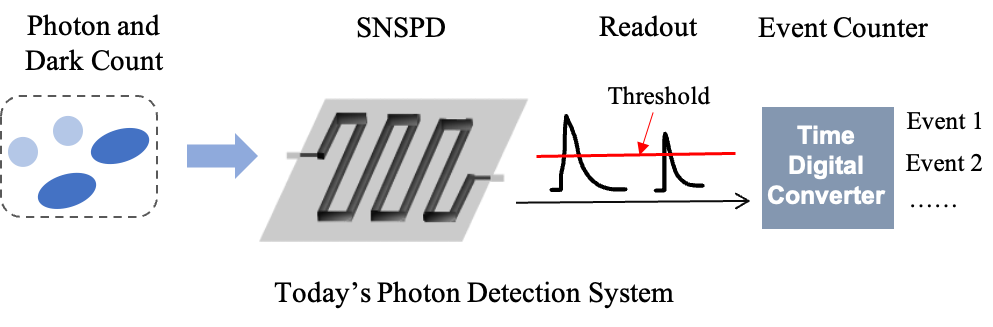}
\caption{Today's single photon detection system. The working principle is binary in nature, utilizing a predefined threshold as a comparator to count the number of detected events regardless of whether they are real photons or dark counts.}

\label{fig:time_tagger}
\vspace{-2.8mm}
\end{figure}

A typical photon detection system today consists of a single photon detector (e.g., SNSPD) connected to a time-digital converter (TDC) device for recording the arrival times and counting the number of each detected event, regardless of whether the event is caused by a real photon or the dark count. Fig.~\ref{fig:time_tagger} illustrates the simplified architecture of today's detection system. In Fig.\ref{fig:time_tagger}, the real photon and dark count from the environment cause the local temporal temperature change of SNSPD, and such a change is converted into the electrical signal readout with fast transit edge and is captured by the time digital converter.  As shown in the Fig.~\ref{fig:time_tagger}, during the process of recording a signal, the time digital converter uses a predefined threshold to compare the SNSPD readout signal, and one event is recorded when the digitized signal is bigger than the predefined threshold. In such scenario, the time digital converter behave a passive event counter with binary nature of detecting the presence of photon via the trigger of threshold. Specifically, today's detection system considers any recorded event as a photon regardless of whether the signal is caused by a real photon or not. And such detection systems have been used on many practical quantum applications, such as quantum key distribution~\cite{terhaar2023ultrafast} and entanglement distribution~\cite{chapman2022hyperentangled} over quantum networks. The system currently does not allow recording and analysis of the full waveforms of the SNSPD output. Passive operation,  and absence of data recording make the current photon detection systems difficult to adapt to identify the dark count and eliminate the corresponding impact on the increasingly complex quantum information processing tasks. 

\vspace{-0.3cm}
\section{System Design}
\label{sec:systemdesign}
This section first presents the hardware platform,  then we describe the signal extraction approach, and conclude with a discussion of the SNSPD circuit.

\subsection{Hardware Platform}

\begin{figure}[h!] 
\centering
\includegraphics[width=0.8\linewidth]{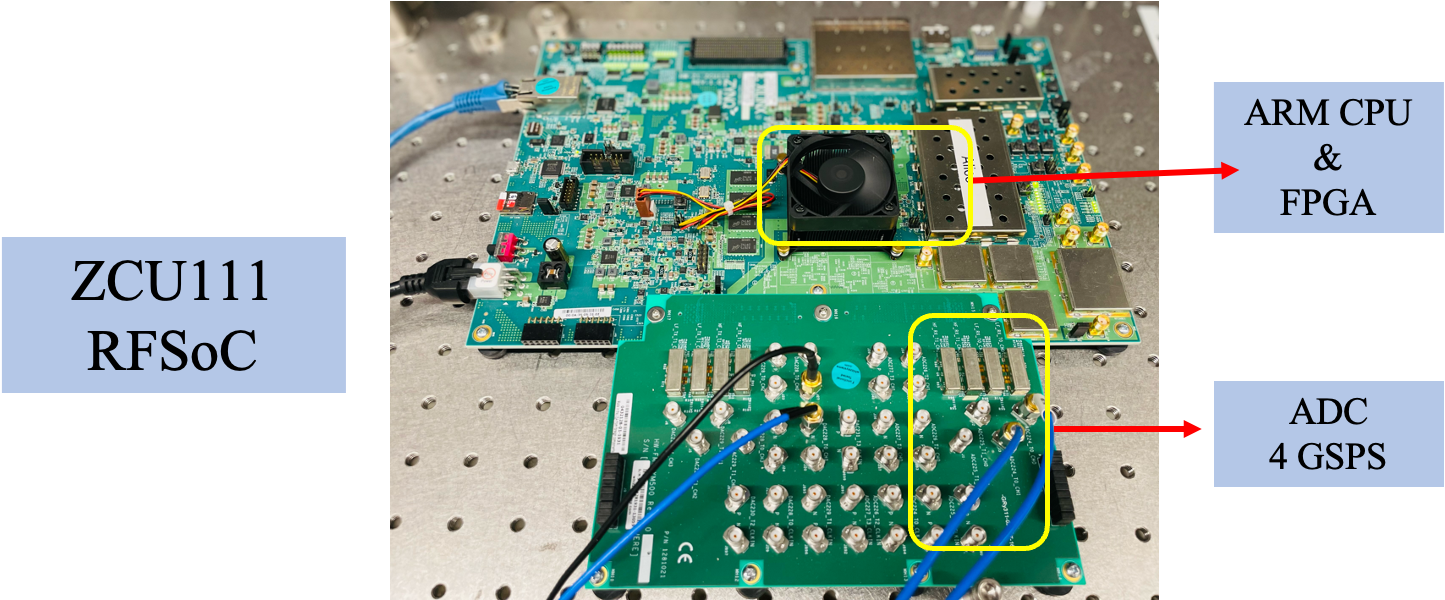}
\caption{Hardware  platform of \name ~based on the ZCU111 RFSoC board. 
The ARM CPU runs a Python-based PYNQ environment to configure the FPGA, which controls the ADCs and processes signals in real time. 
}
\label{fig:zcu111}
\vspace{-0.5cm}
\end{figure}

The hardware platform for \name ~is built on the AMD Xilinx ZCU111 RFSoC evaluation board, shown in Fig.~\ref{fig:zcu111}.  
As highlighted in Fig.~\ref{fig:zcu111}, the board integrates a high-speed ADC subsystem (up to 4GS/s), a programmable logic (FPGA) fabric, 
and an ARM-based processing system on a single system-on-chip (SoC), providing a tightly coupled compute and I/O stack. 
In the \name ~prototype, the PYNQ runtime on the ARM processor orchestrates configuration and control, while the FPGA executes the real-time signal path. The SNSPD readout is continuously sampled by the ADC; each validated event is time-stamped, counted, and its full digitized waveform is captured and saved in on-board memory for downstream processing. Additionally, the architecture on this evaluation board  provides essential advantages for practical fast quantum signal processing, which includes simplified control and configuration, rapid data collection, and low-latency interaction between acquisition and processing stages.

\subsection{Event Drive Signal Extraction}
\begin{figure}[h!] 
\centering
\includegraphics[width=0.8\linewidth]{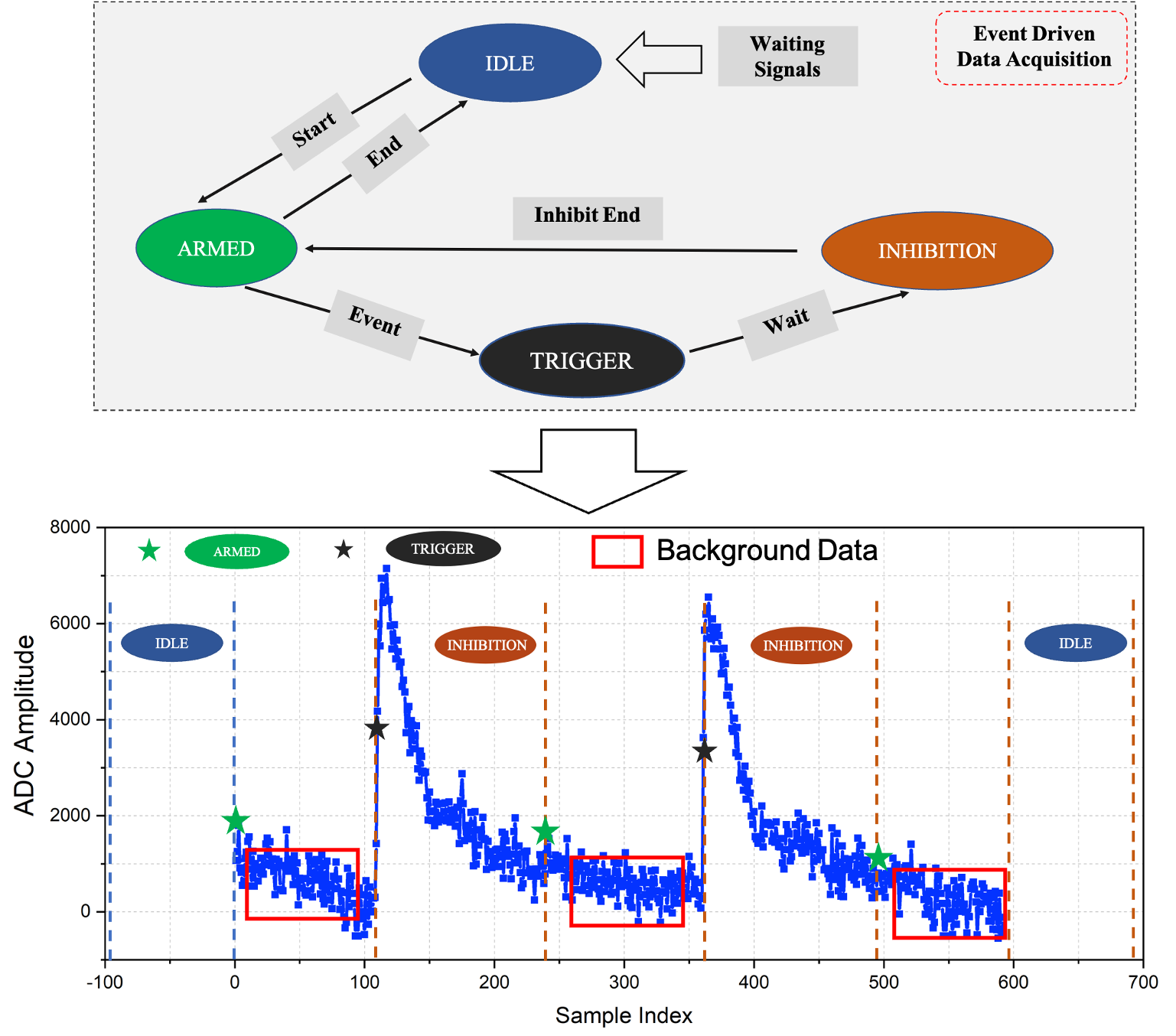}
\caption{Event driven acquisition in \name. 
The system transitions between \texttt{IDLE}, \texttt{ARMED}, \texttt{TRIGGER}, and \texttt{INHIBITION} states to capture valid waveforms only while suppressing continuous background data highlighted as red square .
}
\label{fig:event}
\vspace{-0.3cm}
\end{figure}

This section describes the event-only data acquisition approach. 
\name ~employs an event-driven signal extraction approach, implemented on the open source framework of the QICK~\cite{qick,stefanazzi2022qick}, to capture and save each SNSPD readout. The novelty of this approach is that PhotonIDs captures only valid event-induced signals as needed, while continuously ignoring the idle baseline background data.
The operational mechanism of the proposed approached is demonstrated in  Fig.~\ref{fig:event} with a block diagram and event signal representation. To be simple, this approach includes four state: \texttt{IDLE}, \texttt{AREMD}, \texttt{TRIGGER} and \texttt{INHIBITION} 

At the state of \texttt{IDLE}, the system remains on standby with no active input, waiting for incoming signals. Once a potential signal is detected, the system immediately transitions to the \texttt{ARMED} state (green star marker in Fig.~\ref{fig:event}), which represents the start of active data monitoring.  Once the system is \texttt{AREMD},  the FPGA continuously monitors the ADC data stream, comparing each sampled value against a predefined threshold.
When the first sample exceeds this threshold, one event is recognized, and the system enters the \texttt{TRIGGER} state, labeled as black star mark in the bottom figure of Fig.~\ref{fig:event}. 
Immediately after triggering, the system transitions into the \texttt{INHIBITION} state, during which threshold comparisons are suspended for a fixed waiting period. This prevents multiple triggers from being triggered by the same physical event pulse (rising or long decay tails). The waiting time duration of the \texttt{INHIBITION} sets the post-trigger capture window and therefore the total extracted waveform length.  While the waiting time for the \texttt{INHIBITION} state ends, the system returns to \texttt{ARMED} state to process the next event; and the system falls back to \texttt{IDLE} state if no further signals arrive. 
During this processing loop, the system is designed to save a valid signal with the required length. In this work, we stored a total of 200 samples (8 samples before the triggered point and 192 samples after the triggered point) for each event. This event-driven signal extraction approach ensures that each stored waveform corresponds to a meaningful detection event, significantly reducing memory usage and processing overhead by eliminating continuous background data highlighted as red square.

\subsection{SNSPD Readout Circuit} 
\begin{figure}[h]
\centering
\includegraphics[width=0.8\linewidth]{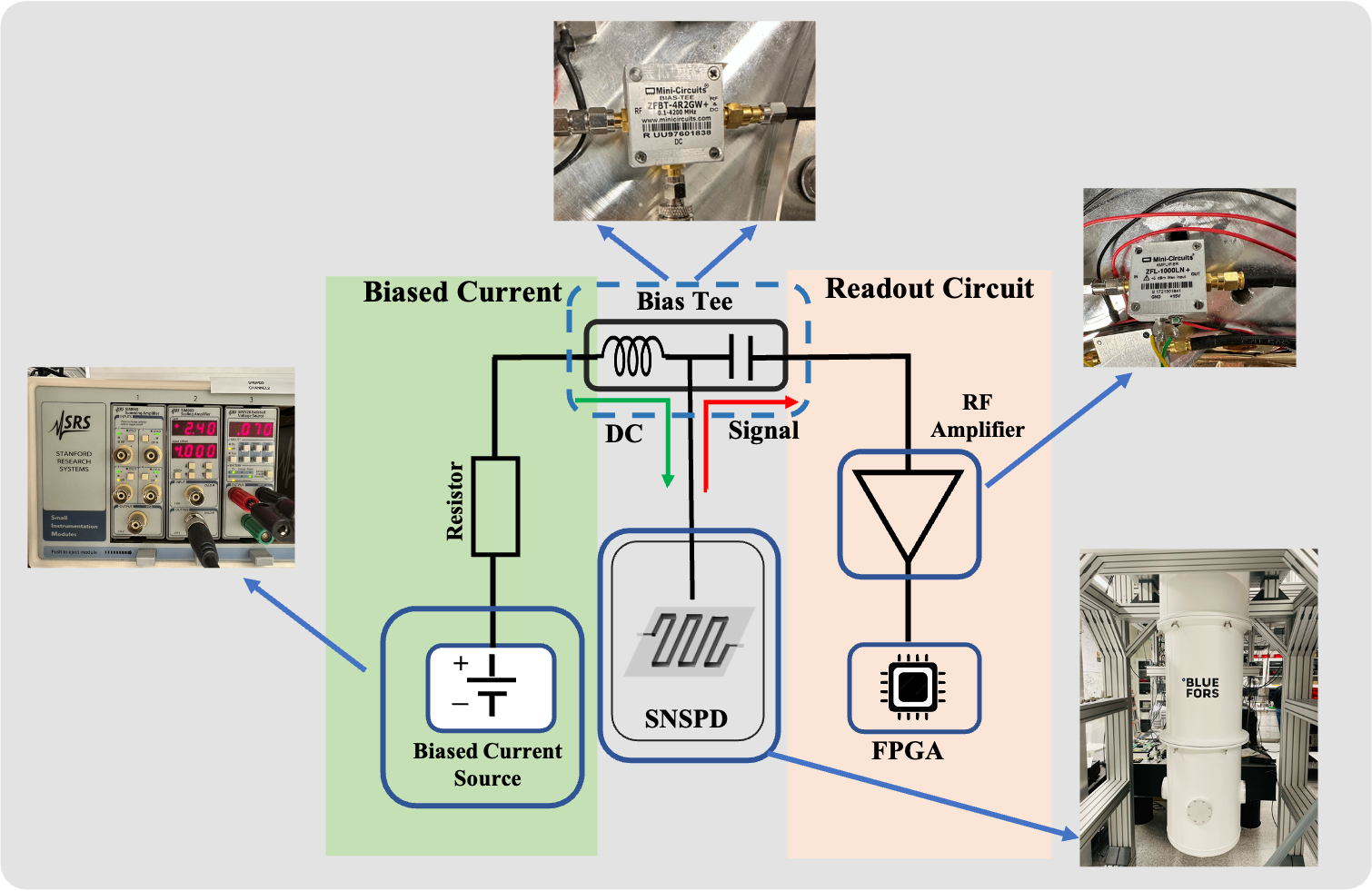}
\caption{SNSPD readout circuit schematics in \name, includes biased current, readout circuit, and the operation environment of SNSPD.}
\label{fig:SNSPD_circuit}
\vspace{-0.8cm}
\end{figure}

This section illustrates the associated SNSPD readout circuit for generating and acquiring
photon or dark count-induced electrical signal variations from the SNSPD, the circuit schematic is shown in Fig.\ref{fig:SNSPD_circuit}. As shown in the figure, the operation of the SNSPD circuit mainly includes three components: the bias current source,  the readout circuit path, and the SNSPD operating environment. In our work, the superconducting nanowire is fabricated with tungsten silicide (WSi)\cite{zhang2016characteristics}, and operates at a helium-based dilution refrigerator (BlueFors SD series)\cite{lake2021bluefors} with approximately 800~millikelvin (mK). At this temperature, the nanowire exhibits zero resistance.

In our work, the biased current is supplied by a Stanford Research Systems SIM980 DC source, providing microampere-level resolution. Under conditions of no absorbed photon or dark count, the bias current circulates entirely through the superconducting nanowire at zero resistance. When the photon or dark count is absorbed, a fast transit signal is generated due to the resistance variation of the nanowire induced by the photon or dark count.

% a localized temperature rise induces a sudden increase in resistance (on the order of kilo-ohms), forcing the bias current to divert through a parallel load resistor. This current redirection produces a detectable signal pulse used to infer the photon’s presence and features.

To separate the DC bias current from the photon/dark count-induced electrical signal, a bias-tee (Mini-Circuits ZFBT-4R2GW+) is used to route the DC path to the SNSPD while allowing the generated electrical signal to pass through to the readout circuit path. The original generated circuit is amplified using a low-noise broadband RF amplifier (Mini-Circuits ZFL-1000LN+), powered by a Siglent SPD3303C power supply. The amplifier boosts the signal amplitude without distorting the temporal profile, ensuring that the waveform's timing characteristics are preserved for digitization and downstream machine learning processing.

% The proposed \name ~system employs an amorphous WSi-based SNSPD~\cite{oripov2023superconducting} with a standard biasing circuit. A cryogenic refrigerator~\cite{bluefors} is used to maintain a temperature of 100 mK.
% % Within this setup, the WSi-SNSPD is positioned as indicated by the red block in Figure\ref{subfig:snsps_dilute}.
% At 100 mK temperature, the WSi nanowire remains in a superconducting state exhibiting near-zero resistance. 
% To bias the detector, an isolated DC voltage source connected in series to a $10 K \Omega$ resistor provides a bias current at the micro-ampere level. The bias current goes through a bias tee to the SNSPD. A quench of the superconducting current in the SNSPD triggered by a photon absorption event diverts the current into a shunt resistor connected in parallel with the SNSPD. This causes a voltage spike that is picked up by RF amplifiers and gets recorded by the FPGA. This raw data is used for training and testing of the neutron network model. The subsequent section will provide a detailed physical modeling of the SNSPD's working mechanism, providing physical intuitions and optimal configurations for the \name ~implementation. %In our work, the voltage source is provided by SRS SIM 928 isolated voltage supply, bias tee is ZFBT-4R2GW, and RF amplifier ZFL-1000LN. 

\section{Theory of Hybrid ML Model}
In this section, we will present the proposed hybrid ML model in \name ~and the corresponding inference pipeline. 
\label{sec:theory}

\vspace{-0.5cm}
\subsection{Motivation of Hybrid Machine Learning Model}

\begin{figure}[h]
\centering
\includegraphics[width=0.8\linewidth]{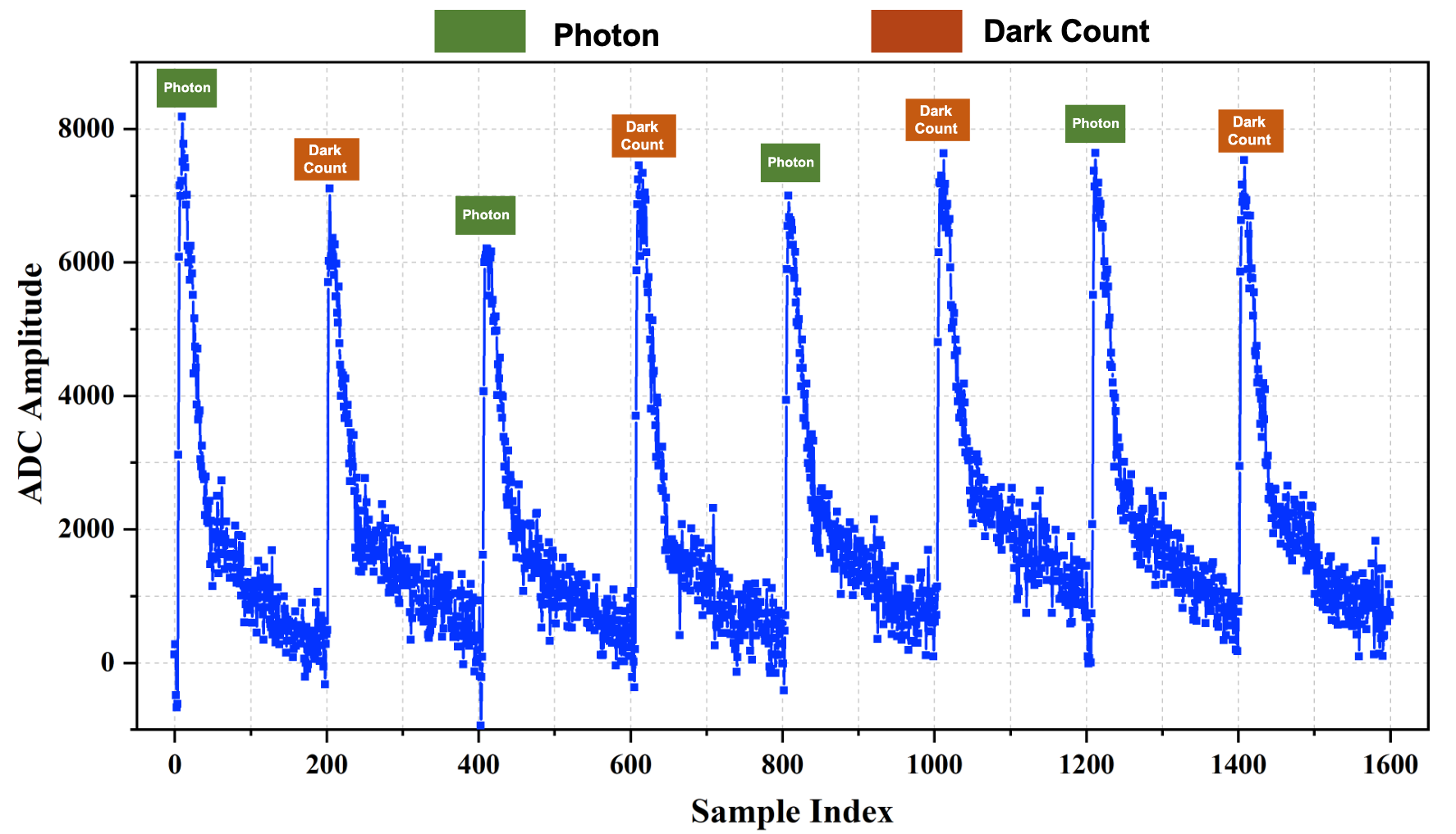}
\caption{Comparison between photon and dark count associated with the captured raw ADC data}
\label{fig:photon_dark_comparison}
\vspace{-0.8cm}
\end{figure}

This work in \name~ considers single-event classification in \name, distinguishing photon and dark count from SNSPD readout waveforms. Fig. \ref{fig:photon_dark_comparison} depicts the representative raw ADC amplitude for both the photon and the dark count. In this work, the ADC amplitude is used for the following machine learning processing without converting the ADC amplitude to voltage.  As shown in Fig. \ref{fig:photon_dark_comparison}, both photon and dark counts exhibit nearly identical pulse morphology, a rapid rising edge followed by a slower decay. Additionally, the distribution of peak amplitude and the temporal width between photon and dark count significantly overlap. These features observed from the raw data make the simple thresholding and template matching ineffective and render naive learning directly on the raw data unreliable.

At the same time, independent studies\cite{holzman2019superconducting}~on SNSPD as the photon detector indicate a strong dependence between absorption location along the nanowire and the variations of readout signal waveform characteristics, mainly including the peak amplitude, temporal information in rising edge, falling edge, and full width half maximum (FWHM). These four scalars are therefore meaningful proxies for a latent ``position" along the nanowire, and explain why photon and dark count signal pulses look similar at a glance yet differ in these position-sensitive summaries. More importantly, the high-frequency noise along with raw data and significant fluctuations in the sampled data introduce extra inaccuracy when extracting these four scalars. 

The above observations motivate the hybrid approach adopted in this work, which first stabilizes and smooths the waveform via smoothing and interpolation and then introduces a physics-anchored pseudo position embedding before discriminative classification, as follows:

\begin{itemize}
\item \textbf{Preprocessing for per-event robustness.} Each waveform is denoised by a Savitzky-Golay filter to suppress high-frequency noise while preserving signal shape, then resampled by interpolation to achieve a uniform temporal grid. This combination stabilizes both amplitude and temporal information.
\item \textbf{Physics-anchored pseudo-position (offline ruler).} From the four scalars, we define a dimensionless pseudo-position via kernel density estimates (KDE) given a large amount of collected waveform data analysis, which establishes a consistent ruler tied to the empirical feature distribution and provides an interpretable target scale.
\item \textbf{Single-event regression and calibration (online predictor).} 
As the physics-anchored pseudo position relies on large-scale data analysis and is not available for individual waveform analysis, as targeted by \name. 
A compact 1-D CNN regresses the pseudo-position directly from each full waveform, capturing morphology beyond the four scalars KDE-derived position as supervision. 
A monotone, shape-preserving calibrator (PCHIP) then aligns the raw CNN outputs to the KDE-anchored scale, removing global bias/scale while preserving rank. 
This ``offline anchor, online predictor" design closes the loop between physics-derived summaries and real-time per-event inference.
\item \textbf{Hybrid discrimination.} The calibrated pseudo-position is concatenated with the four scalars and fed to a shallow fully connected classifier for photon and dark count classification decisions, leveraging both physics-aware structure and learned temporal representation.
\end{itemize}

The following section provides the theoretical analysis of the proposed hybrid machine learning model in this work.

\subsection{Proposed Hybrid Machine Learning Model}

\begin{figure*}[t]
    \centering
    \includegraphics[width=\linewidth]{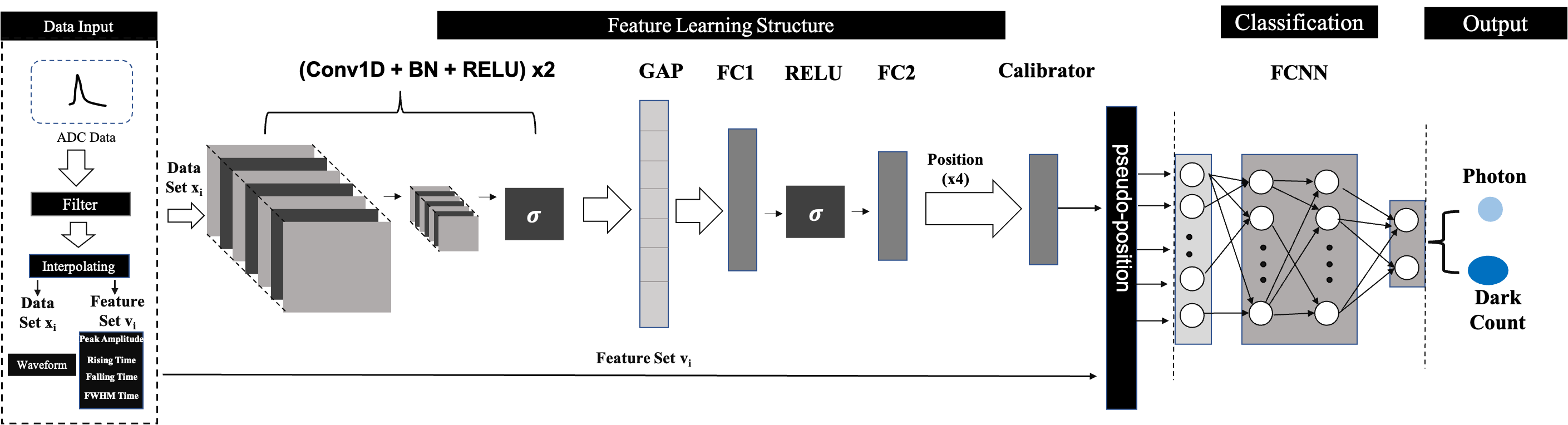}
    \caption{Architecture of the proposed hybrid learning pipeline. 
    \textbf{Left} \emph{Data input \& preprocessing}: the ADC data captured from SNSPD readout waveform is denoised (Savitzky Golay) and interpolated; four scalar features $v=[\text{peak amplitude},\text{rising time},\text{falling time},\text{FWHM time}]$ are extracted. 
    \textbf{Middle} \emph{Feature learning}: a compact 1-D CNN (Conv+BN+ReLU blocks with global average pooling and two fully connected layers) produces a temporal embedding and regresses a four-dimensional pseudo position $\tilde{\mathbf p}=f_\theta(x)$. A calibrator PCHIP maps $\tilde{\mathbf p}$ to calibrated positions $\hat{\mathbf p}=\mathbf g(\tilde{\mathbf p})$ (omitted in the diagram for clarity). 
\textbf{Right} \emph{Classification}: the calibrated positions $\hat{\mathbf p}$ are concatenated with $\mathbf v$ and passed to a shallow FCNN; a softmax outputs the posterior over classes (photon or dark count). 
% The design preserves physically interpretable intermediates while enabling fast, end-to-end inference.
}
    \label{fig:cnn_mlp}
\end{figure*}

% {\color{teal}
% The core of \name~is a hybrid machine learning model designed to accurately discriminate between photon and dark count events based on their waveform signatures.
Fig.~\ref{fig:cnn_mlp} demonstrates the main architecture of the proposed hybrid machine learning model.
% As illustrated in Fig.~\ref{fig:cnn_mlp}, this model integrates scalar feature extraction with deep learning in a multi-stage pipeline. 
Generally, the model consists of: (1) data input and preprocessing, (2) feature learning structure with a 1D CNN regressor and calibrator to get the ``pseudo-position", and (3) a final fully connected neural network (FCNN) for photon and dark count classification. 
% This architecture is designed to produce physically interpretable intermediate features, ensuring both high accuracy and model transparency.

\subsubsection{Data Input and Processing}
\label{subsubsec:data}
Each readout signal from the SNSPD is captured as a discrete time-series waveform, denoted as a vector $\mathbf{x}_i^{raw} \in \mathbb{R}^L$, where $i$ is the event index and $L$ is the number of temporal sampling points. After applying Savitzky-Golay filtering and interpolation for denoising and up-sampling, we extract a set of four scalar features from each preprocessed waveform $\mathbf{x}_i$. These features, which provide a low-dimensional summary of the waveform's morphology, are: peak amplitude, rising time, falling time, and FWHM time. We represent these features as a vector:
$$
\mathbf{v}_i = [v_{i, \text{peak}}, v_{i, \text{rise}}, v_{i, \text{fall}}, v_{i, \text{fwhm}}]^T \in \mathbb{R}^4
$$
This vector is computed via a feature extraction function, such that $\mathbf{v}_i = \text{feature\_extraction}\,(\mathbf{x}_i)$.

\subsubsection{KDE-Anchored Position Encoding}
\label{subsubsec:kde}
To create a robust and physically grounded target for our regression model, we transform the handcrafted feature space into a ``pseudo-position" space. This is achieved by anchoring each feature to its statistical distribution across the training dataset. For each of the four feature types, indexed by $k \in \{\text{peak, rise, fall, fwhm}\}$, we first estimate an empirical probability density function $\hat{f}_k(t)$ using a Gaussian Kernel Density Estimator (KDE):
\begin{equation}
\hat{f}_k(t) = \frac{1}{N h_k} \sum_{i=1}^{N} \frac{1}{\sqrt{2\pi}} \exp\left(-\frac{1}{2}\left(\frac{t - v_{i,k}}{h_k}\right)^2\right)
\label{eq:kde}
\end{equation}
where $\{v_{i,k}\}_{i=1}^N$ are the observed feature values from the $N$ training samples and $h_k$ is the kernel bandwidth, \(t\) is the continuous evaluation variable on the same one-dimensional feature axis as \(v_{i,k}\). We define the mode of this distribution as $\mu_k = \arg\max_t \hat{f}_k(t)$. A dimensionless position ruler, $\Delta_k$, is then established to normalize the feature's dynamic range:
$$
\Delta_k = \frac{\max_i(v_{i,k}) - \min_i(v_{i,k})}{F}
$$
where $F$ is a constant scaling factor. Using this, we define the ground-truth pseudo-position for the $k$-th feature of the $i$-th event as:
\begin{equation}
p_{i,k} = \frac{v_{i,k} - \mu_k}{\Delta_k}
\label{eq:pos_encoding}
\end{equation}
This process yields a 4-dimensional ground-truth position vector $\mathbf{p}_i = [p_{i,1}, p_{i,2}, p_{i,3}, p_{i,4}]^T$ for each waveform, which serves as the target for the CNN regression task.

\subsubsection{CNN Regression of Pseudo-Positions}
\label{subsubsec:regress}
We employ a lightweight 1D CNN to learn a mapping $f_\theta: \mathbb{R}^L \to \mathbb{R}^4$ that directly regresses the pseudo-position vector from the preprocessed input waveform. The network, with parameters $\theta$, predicts a raw position vector $\tilde{\mathbf{p}}_i = f_\theta(\mathbf{x}_i)$. The core of the network consists of 1D convolutional layers where the activation $z_{c,l}$ for the $c$-th channel at position $l$ is computed as:
\begin{equation}
z_{c,l} = \sigma\left(b_c + \sum_{c'=1}^{C_{in}} \sum_{j=0}^{K-1} W_{c,c',j} u_{c',l+j}\right)
\label{eq:conv1d}
\end{equation}
Here, $\mathbf{u}$ is the input feature map from the previous layer, $K$ is the kernel size, $\mathbf{W}$ are the convolutional weights, $\mathbf{b}$ is the bias vector, and $\sigma(\cdot)$ is the ReLU activation function. A global average pooling (GAP) layer followed by two fully connected layers) then maps the learned features to the four pseudo-position outputs. The network is trained by minimizing the mean squared error (MSE) between the predicted positions and the KDE-anchored ground-truth positions.
% \begin{equation}
% \mathcal{L}_{\text{MSE}}(\theta) = \frac{1}{N} \sum_{i=1}^{N} \| f_\theta(\mathbf{x}_i) - \mathbf{p}_i \|_2^2
% \label{eq:mse_loss}
% \end{equation}

\subsubsection{Monotone Shape-Preserving Calibration}
\label{subsubsec:monotone}
The raw position predictions $\tilde{\mathbf{p}}_i$ from the CNN often exhibit systematic affine bias (i.e., offset and scaling errors) relative to the ground-truth targets $\mathbf{p}_i$. To correct this, we introduce a non-parametric calibration step. For each of the four scalar features, we learn a monotone, shape-preserving mapping $g (\cdot)$ using a piecewise cubic Hermite interpolating polynomial (PCHIP). This function is trained on pairs of raw predictions and ground-truth values from the training set. At inference time, this function maps the 1D CNN-regressed raw position to a calibrated one. This calibration is applied to produce a calibrated position vector:
\begin{equation}
\hat{\mathbf{p}}_i = g(\tilde{\mathbf{p}}_i) = [g_1(\tilde{p}_{i,1}), g_2(\tilde{p}_{i,2}), g_3(\tilde{p}_{i,3}), g_4(\tilde{p}_{i,4})]^T
\label{eq:calibration}
\end{equation}
The PCHIP method ensures that the calibration is monotonic and does not introduce spurious oscillations, thereby preserving the rank information learned by the CNN while correcting for systematic bias.

\subsubsection{FCNN for Final Classification}
\label{subsubsec:classification}
Finally, a discriminative FCNN performs the classification between photon and dark count events. The input to this classifier is a hybrid feature vector $\mathbf{z}_i \in \mathbb{R}^8$, formed by concatenating the original handcrafted scalar features $\mathbf{v}_i$ with the \textit{calibrated} pseudo-position vector $\hat{\mathbf{p}}_i$:
\begin{equation}
\mathbf{z}_i = [\mathbf{v}_i, \hat{\mathbf{p}}_i]^T = [v_{i,\text{peak}}, \dots, v_{i,\text{fwhm}}, \hat{p}_{i,1}, \dots, \hat{p}_{i,4}]^T
\label{eq:hybrid_feature}
\end{equation}
This combined feature vector is passed through a shallow FCNN, denoted $f_{\text{cls}}(\cdot)$, followed by a softmax activation function to produce the final class probabilities. For a set of $K$ classes (here, $K=2$), the predicted probability vector is:
$$
\hat{\mathbf{y}}_i = \text{softmax}(f_{\text{cls}}(\mathbf{z}_i)) \in \mathbb{R}^K
$$
The final classification is given by the class with the highest probability. This hierarchical design, summarized in Algorithm \ref{alg:pipeline}, ensures that the model leverages both low-level waveform morphology and high-level, physically-anchored learned features, leading to robust and interpretable discrimination.

\begin{algorithm}[h]
\SetAlgoLined
\KwIn{SNSPD waveform $\mathbf{x}_i^{raw} \in \mathbb{R}^L$}
\KwOut{Predicted class $\hat{y}_i \in \{0 \text{ (dark)}, 1 \text{ (photon)}\}$}
\ForEach{incoming waveform $\mathbf{x}_i^{raw}$}{
    \tcp{(1) Preprocess waveform}
    Apply Savitzky-Golay denoising and time interpolation on $\mathbf{x}_i$\;
    \tcp{(2) Handcrafted scalar features}
    Compute $\mathbf{v}_i = \text{feature\_extraction}(\mathbf{x}_i) \in \mathbb{R}^4$\;
    \tcp{(3) CNN regression of raw positions}
    $\tilde{\mathbf{p}}_i = \text{CNN}(\mathbf{x}_i) \in \mathbb{R}^4$\;
    \tcp{(4) Monotone calibration (per scalar feature)}
    $\hat{\mathbf{p}}_i = g(\tilde{\mathbf{p}}_i) = [g_1(\tilde{p}_{i,1}), \dots, g_4(\tilde{p}_{i,4})]^T$\;
    \tcp{(5) Hybrid fusion + FCNN}
    Concatenate hybrid feature $\mathbf{z}_i = [\mathbf{v}_i, \hat{\mathbf{p}}_i]^T \in \mathbb{R}^8$\;
    Classify event: $\hat{y}_i = \text{FCNN}(\mathbf{z}_i)$\;
}
\Return {Predicted class labels $\{\hat{y}_i\}$}
\caption{Hybrid ML Inference Pipeline}
\label{alg:pipeline}
\end{algorithm}

% {\color{teal}
\subsection{Inference Pipeline}
The end-to-end inference pipeline is a multi-stage process designed to transform a raw SNSPD waveform into a final class label (photon or dark count). This procedure, detailed in Algorithm~\ref{alg:pipeline}, ensures interpretability and high performance. The pipeline proceeds as follows:

\begin{enumerate}
    \item \textbf{KDE-Anchored Position Embedding:} For each handcrafted feature vector $\mathbf{v}_k$, a corresponding pseudo position vector $\mathbf{p}_k$ is established using kernel density estimation (KDE). This provides a stable, physically meaningful regression target.

    \item \textbf{CNN-Based Pseudo-Position Regression:} A lightweight 1D CNN, $f_{\theta}$, takes the preprocessed waveform $\mathbf{x_{i}}$ as input and directly regresses a raw 4-dimensional pseudo-position vector $\tilde{\mathbf{p}} = f_{\theta}(\mathbf{x_{i}})$.

    \item \textbf{Monotonic Calibration:} A shape-preserving monotonic calibrator, $g$, maps the raw pseudo-positions to calibrated positions, $\hat{\mathbf{p}} = g(\tilde{\mathbf{p}})$. This non-parametric step corrects for global bias and scaling errors from the CNN while preserving the learned ordinal information.

    \item \textbf{Hybrid Classification:} A shallow FCNN performs the final classification. Its input is a hybrid feature vector $\mathbf{z_i} = [\mathbf{v}_i, \hat{\mathbf{p}}_i]^T$, which concatenates the original handcrafted features $\mathbf{v}_i$ with the calibrated pseudo-position vector $\hat{\mathbf{p}}_i$.
\end{enumerate}

This hierarchical design maintains physically interpretable intermediate quantities and employs monotonic calibration to control for overfitting, thereby achieving accurate classification with sub-millisecond inference latency.

\section{Result analysis and Evaluation}
\label{sec:experimental_evaluation}
This section evaluates the performance of the proposed hybrid machine learning model in classifying photon and dark count. We begin with data collection and the associated preprocessing. Then, the experimental configuration and the evaluation metric are presented. Model performance, pseudo-position, classification accuracy, and t-SNE visualization are demonstrated

\subsection{Dataset Collection Setting}
\begin{figure}[h]
\centering
\includegraphics[width=0.9\linewidth]{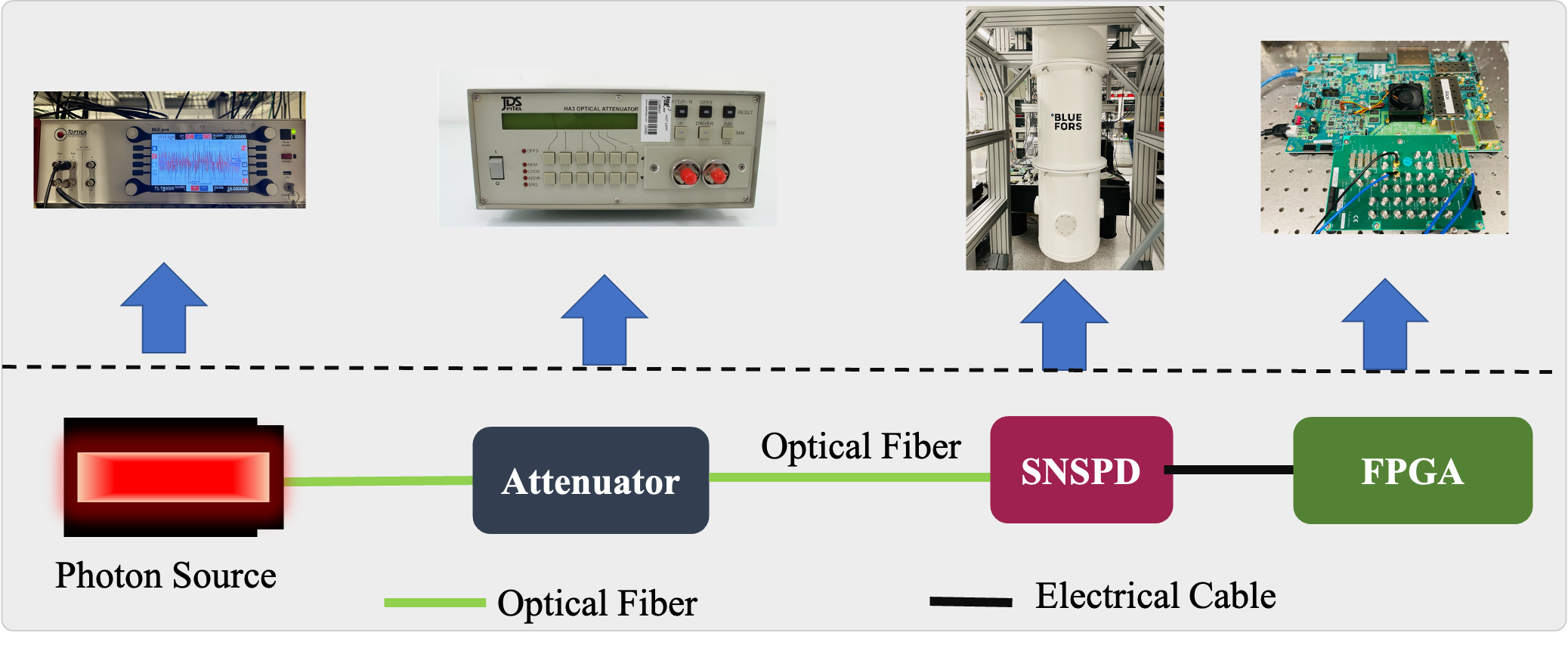}
\caption{Experimental Setup for the data collection}
\label{fig:photon_dark_collection}
\vspace{-0.4cm}
\end{figure}

This section describes the experimental setup for data collection and associated data preprocessing used in the proposed \name~ system.  Fig.~\ref{fig:photon_dark_collection} illustrates the experimental setup for the photon data and dark count data collection. The experimental setup includes the photon source provided by TopTica DLC Pro diode laser source; tunable optical attenuator to adjust the photon rate to avoid the SNSPD saturation; dilute fridge produced by BlueFors housed the SNSPD with operation temperature of millikelvin; and ZCU 111 evaluation board to capture the complete waveform of each SNSPD readout signal based on the proposed data acquisition approach mentioned in Section III.  

% The waveform dataset was acquired using a designed high-speed data acquisition system implemented on an FPGA. The system incorporates a high-resolution analog-to-digital converter (ADC) with a sampling rate of 2GSPS, enabling precise temporal capture of photon induced signal from the SNSPD readout circuit. Notably, the FPGA logic is designed to suppress background noise by triggering only on valid photon events, thereby ensuring that the recorded waveforms correspond exclusively to photon-induced signals. 

% While proceeding the data collection with the experimental setup as shown in Fig.\ref{fig:experiment_setting}, all ambient light sources were eliminated to minimize environmental noise, such as room light.

To build clean training and testing data, we acquired photon-only and dark count-only waveforms across multiple non-consecutive sessions (different hours and days) in the same underground laboratory. For the photon dataset, the experiment was performed in a dark laboratory and a 3-meter single-mode fiber was coupled between the photon source and the SNSPD to suppress ambient interference. A TopTica DLC Pro diode laser was tuned to 1535nm (to align with the erbium emission case study) and attenuated with a calibrated variable optical attenuator to yield a photon rate of 8k events/s at the detector input. At this operating point, the accidental noise output is negligible for the downstream machine learning analysis.  During acquisition, we continuously monitored the count rate and saved full waveforms using the event-driven trigger on the FPGA; only validated events were buffered to memory. In total, we collect $\approx 200000$ labeled waveforms for photon and dark count separately. These data would be used for the following hybrid ML analysis. 

These procedures (dark room operations, short fiber, optical attenuation, event-driven data capture, multiple non-consecutive sessions (different hours and days)) ensure that the labels are reliable and the dataset is sufficiently large and representative for the hybrid ML study.

\vspace{-0.5cm}
\subsection{Dataset Preprocessing}

Each detection event generates a one-dimensional waveform sampled over a 100ns time window. In subsequent processing, a shorter segment of the waveform is extracted based on the temporal profile of the photon response, ensuring focus on the most informative portion of the signal.  To suppress high-frequency electrical noise and preserve the signal clarity, a Savitzky Golay filter (window size = 11, polynomial order = 3) was applied to each raw waveform. The filtered waveforms were then interpolated using interpolation with a 20$\times$ up-sampling factor, increasing 20 times of the number of data points per waveform. This high-resolution resampling enhances the signal smoothness and temporal precision, facilitating improved alignment and more robust feature extraction for learning. 

\begin{figure}[h!]
\vspace{-6mm}
\centering
\subfloat[]{
\includegraphics[width=0.45\linewidth]{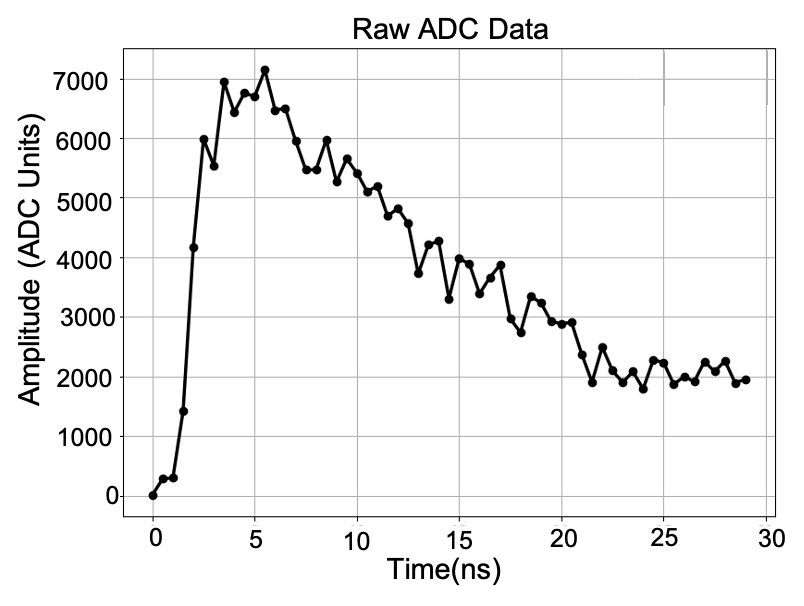}}
\subfloat[]{
\includegraphics[width=0.45\linewidth]{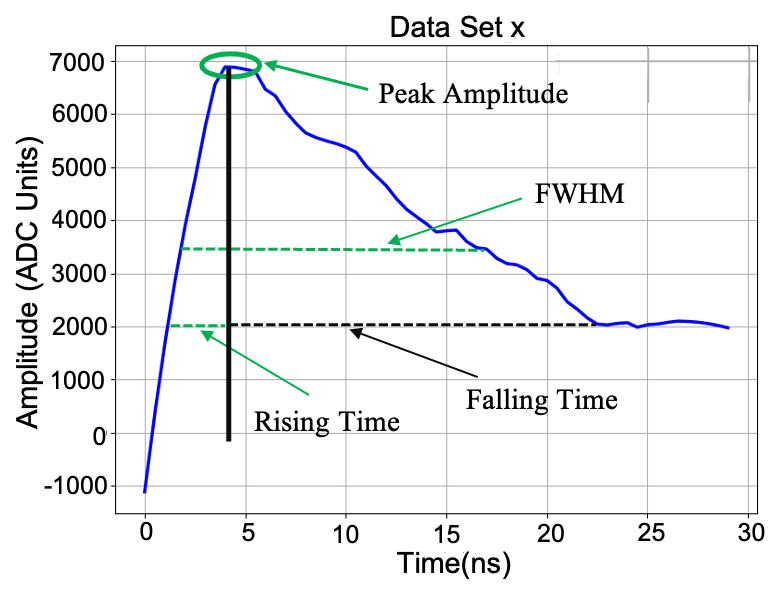}}
\caption{Representative SNSPD waveforms before (a) and after (b) preprocessing. 
The preprocessing pipeline includes Savitzky Golay filtering and 20 times cubic interpolation to enhance temporal resolution. Peak amplitude, rising time, falling time, and FWHM are static features extracted from the processed signal.}
\label{fig:data_smoothing}
\vspace{-6mm}
\end{figure}

Fig.~\ref{fig:data_smoothing} visualizes a typical waveform before and after preprocessing within a representative 30~ns window, sufficient to highlight waveform shape differences across different wavelengths. Each final waveform is represented as a 20 times temporal resolution vector, and four statistical features, peak amplitude, rising time, falling time, and FWHM time, are extracted per waveform to support the hybrid ML model.

\subsection{Experimental Configuration}
All experiments were run on a local workstation (Intel Core i9 CPU, 64\,GB RAM, Windows~11). 
Each preprocessed waveform is processed by a compact 1-D CNN that predicts four standardized pseudo-positions.
The network comprises two convolutional blocks: a Conv1D with 1 input and 64 output channels (kernel size 3, padding 1) followed by batch normalization and ReLU, and a second Conv1D with 64 output channels reduced to 32 (kernel size 3, padding 1) again followed by batch normalization and ReLU. A global average pooling over the temporal dimension reduces the feature map to a 32-dimensional vector.
The pooled features are then passed through a fully connected layer (32 to 128) with ReLU and dropout ($p{=}0.2$), and a final linear layer (128 to 4) produces the four outputs.
We train with mean-squared error with learning rate $5\times 10^{-4}$ and batch size $64$ for $50$ epochs. The FCNN classifiers are trained up to 300 epochs(learning rate of 0.0001 and batch size of 32) with cross-entropy loss. The FCNN classifier consists of four fully connected layers with 256, 128, 64, and 32 neurons, respectively, followed by ReLU activations and a final softmax output layer for two-class classification.

{\color{black}
\vspace{-0.5cm}
\subsection{Evaluation Metrics}
The performance of the \name~model is evaluated at its two primary stages: the intermediate pseudo-position regression and the final photon/dark count classification.

For the regression stage, we report the mean absolute error (MAE), $MAE = \frac{1}{N}\sum_{i}|\hat{p}_{i,k}-p_{i,k}|$ root mean squared error (RMSE), $RMSE = \sqrt{\frac{1}{N}\sum_{i}(\hat{p}_{i,k}-p_{i,k})^{2}}$; Coefficient of determination ($R^2$), $R^{2} = 1-\frac{\sum_{i}(\hat{p}_{i,k}-p_{i,k})^{2}}{\sum_{i}(p_{i,k}-\overline{p}_{k})^{2}}$; and tolerance accuracy, defined as $\frac{1}{N}\sum_{i}\mathbb{I}\{|\hat{p}_{i,k}-p_{i,k}|\le\tau\}$, where $\mathbb{I}\{\cdot\}$ is the indicator function.

For the final classification task, we report overall accuracy, per-class precision, recall, F1-score, and the receiver operating characteristic area under the curve (ROC-AUC) to assess the model's discriminative power.
}
\subsection{Pseudo Position Regression and Calibration Results}

Built on the KDE anchored position embeddings for four scalar features, this section statistically demonstrates the performance comparison of pseudo position with the proposed 1D CNN regression and the calibration. 

\begin{figure*}[t]
    \centering
    \includegraphics[width=\linewidth]{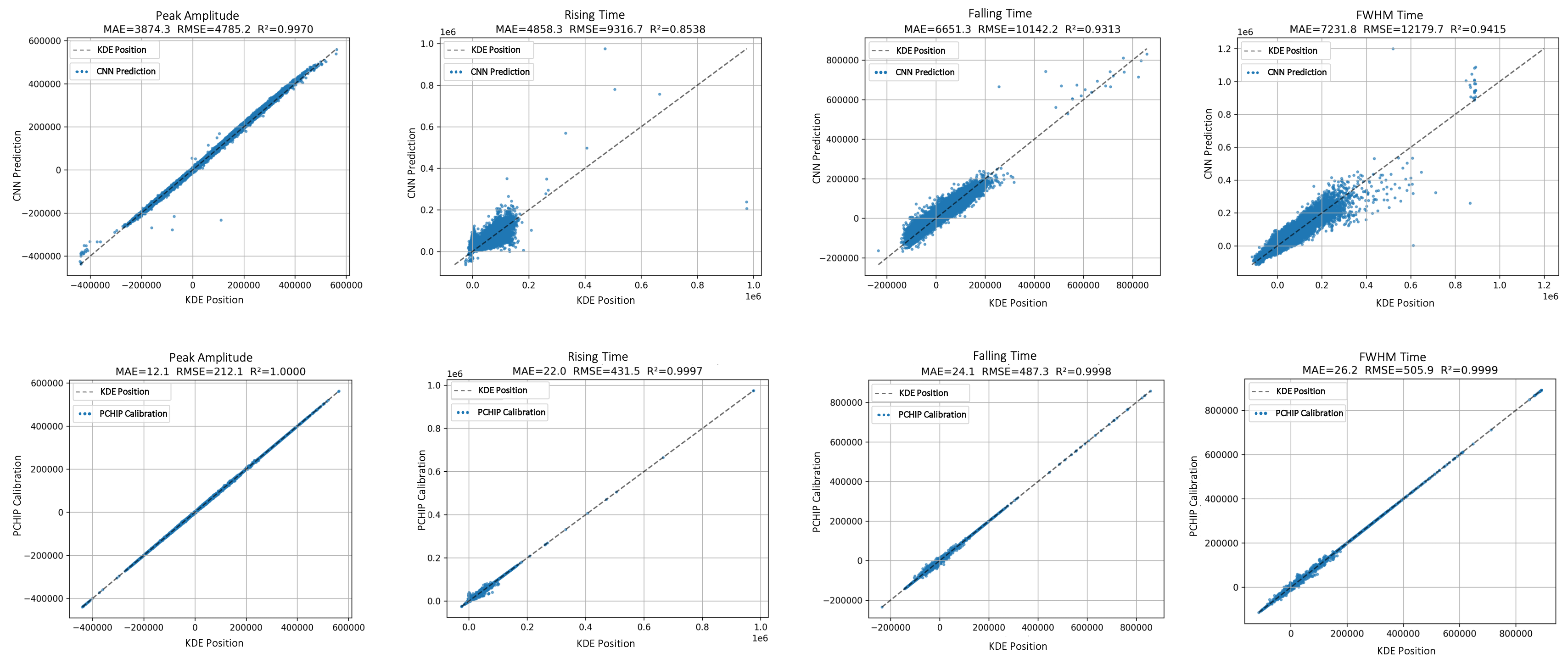}
    \caption{Pseudo-position comparison between regression and calibration. 
  Top row: 1D CNN predictions vs.\ KDE-anchored positions for (a) Peak Amplitude, (b) Rising Time, (c) Falling Time, and (d) FWHM Time. 
  Bottom row: the same targets after applying the PCHIP calibrator to the CNN outputs.  
  Calibration removes global bias and slope mismatch in the raw CNN outputs, yielding near-identity mappings and large error reductions (Table~\ref{tab:ablation_position}).}
    \label{fig:cnn_calib}
    \vspace{-0.4cm}
\end{figure*}

% Built on the KDE anchored position encoding of Sec.~\ref{sec:theory}, each record provides four physically interpretable targets $p_k$ (Peak Amplitude, Rising Time, Falling, FWHM Time). A compact 1D CNN regresses raw pseudo-positions $\tilde{\mathbf p}=f_\theta(x)$ from the full waveform, and a channel wise monotone, shape-preserving calibrator $\mathbf g$ (PCHIP; Sec .~\ref {sec:theory}) maps them to $\hat{\mathbf p}=\mathbf g(\tilde{\mathbf p})$ as mentioned in Sec .~\ref {sec:theory}.

Fig.~\ref{fig:cnn_calib} visualizes the results. The \emph{top row} shows \emph{raw CNN outputs} versus KDE-anchored positions; the dashed line is the identity. The raw estimates already follow a strong linear trend but exhibit a global affine bias and mild slope distortion, most visible for the parameters: peak amplitude, rising time, falling time, and FWHM time. The \emph{bottom row} plots \emph{calibrated outputs} versus the KDE-anchored positions; after applying $g (\cdot)$ the point clouds collapse onto KDE-anchored positions curve with negligible curvature, indicating that monotone calibration removes bias scale while preserving rank information. 

Quantitatively, calibration yields large error reductions and $R^2$ close to unity, and 
Table~\ref{table:pseudo position} summarizes the improvement. Across all four targets, calibration reduces MAE by $\approx 99.6\text{~to~}99.7\%$, RMSE by $\approx 95.2\text{~to~}95.8\%$, and pushes $R^2$ to $\ge 0.9997$. These gains are shows the strong evidence that the PCHIP map corrects the affine distortion without oscillation, converting systematic error into near-zero residual noise.

\begin{table*}[b]
\centering
\caption{Pseudo-position regression: raw CNN vs.\ calibrated (PCHIP) against KDE-anchored positions. 
$\Delta$ denotes relative error reduction.}
\label{tab:ablation_position}
\small
\setlength{\tabcolsep}{6pt}      
\renewcommand{\arraystretch}{1.25} 
\begin{tabular}{
  l
  S[table-format=4.1]  % MAE-CNN
  S[table-format=4.1]  % MAE-PCHIP
  S[table-format=2.1]  % dMAE%
  S[table-format=5.1]  % RMSE-CNN
  S[table-format=4.1]  % RMSE-PCHIP
  S[table-format=2.1]  % dRMSE%
  S[table-format=1.4]  % R2-CNN
  S[table-format=1.4]  % R2-PCHIP
}
\toprule
\multirow{2}{*}{Target} &
\multicolumn{3}{c}{MAE} &
\multicolumn{3}{c}{RMSE} &
\multicolumn{2}{c}{$R^{2}$} \\
\cmidrule(lr){2-4} \cmidrule(lr){5-7} \cmidrule(lr){8-9}
& {CNN} & {PCHIP} & {$\Delta$MAE (\%)} & {CNN} & {PCHIP} & {$\Delta$RMSE (\%)} & {CNN} & {PCHIP} \\
\midrule
Peak Amplitude & 3847.3 & 12.1 & 99.7 & 4785.2  & 212.1 & 95.6 & 0.9970 & 1.0000 \\
Rising Time    & 4858.3 & 22.0 & 99.7 & 9316.7  & 431.5 & 95.4 & 0.8538 & 0.9997 \\
Falling Time   & 6651.3 & 24.1 & 99.6 & 10142.2 & 487.3 & 95.2 & 0.9313 & 0.9998 \\
FWHM Time      & 7231.8 & 26.2 & 99.6 & 12179.7 & 505.9 & 95.8 & 0.9415 & 0.9999 \\
\bottomrule
\end{tabular}
\label{table:pseudo position}
\end{table*}

In the remainder, we use the calibrated positions $\hat{\mathbf p_i}$, concatenated with the four scalar features $\mathbf v_i$, as inputs to the FCNN classifier. The next subsections analyze confusion matrices, overall classification accuracy, and the separability of the calibrated feature space (t-SNE).

\subsection{Classification Performance Evaluation}
\begin{figure*}[t]
    \centering
    \includegraphics[width=\linewidth]{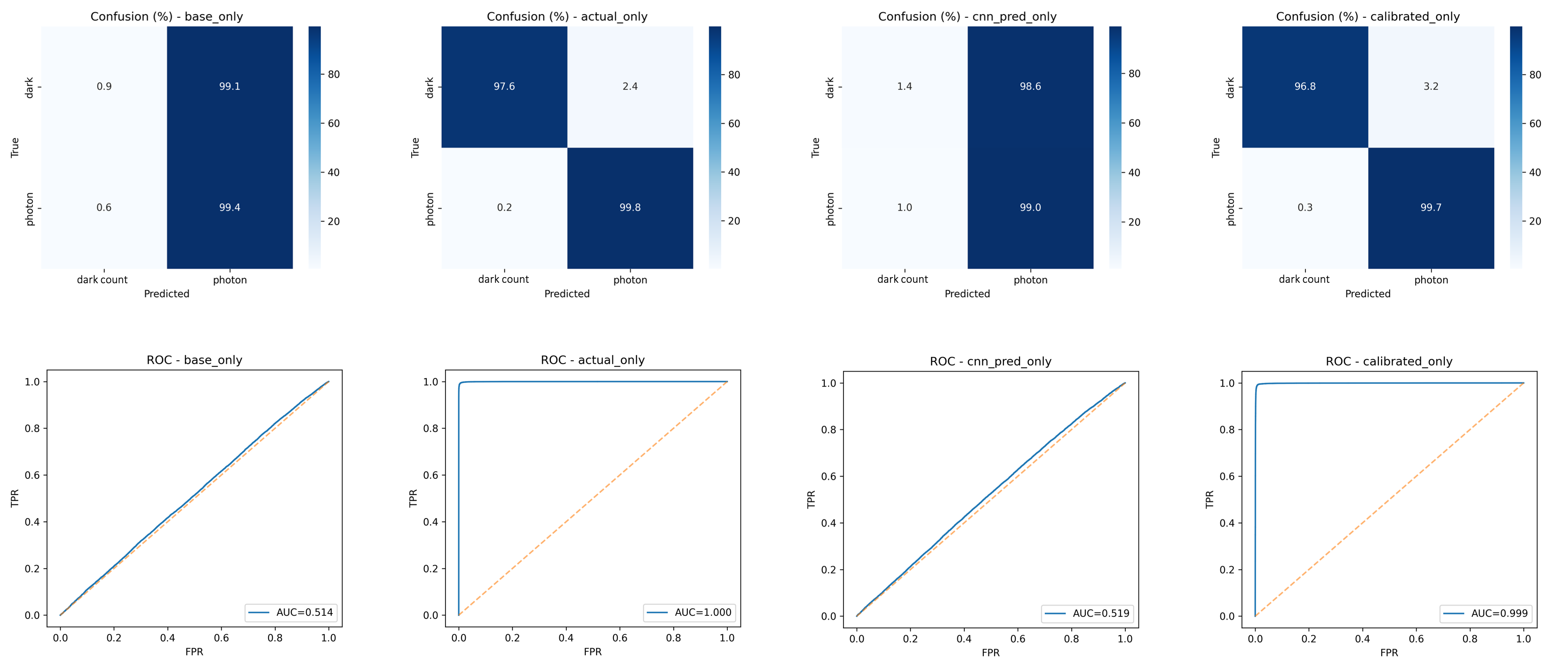}
    \caption{\textbf{Confusion matrices (top) and ROC curves (bottom) for four configurations.}
  Columns from left to right: \emph{base\_only}, \emph{actual\_only}, \emph{cnn\_pred\_only}, and \emph{calibrated\_only}.
  Each confusion matrix is shown in percentage form (row-normalized).
  The ROC curves summarize threshold-free separability; the displayed AUC values (in the legends) correspond to the same test split. 
}
    \label{fig:fcnn_matrix}
\vspace{-0.3cm}
\end{figure*}

This section demonstrates the performance of photon and dark count classification with the scalar features and the calibrated pseudo-position via CNN and PCHIP as the combined input features for the FCNN-based classifier.  To assess how the pseudo-position contributes to separability at the classifier, four input configurations are evaluated under the \emph{same} FCNN, data split, and preprocessing: (i) \texttt{base\_only} four scalar features only; (ii) \texttt{actual\_only}  KDE anchored positions concatenated with the four scalars; (iii) \texttt{cnn\_pred\_only} CNN regressed positions (before calibration) with the four scalars; and (iv) \texttt{calibrated\_only}  PCHIP calibrated positions with the four scalars.  
Fig.~\ref{fig:fcnn_matrix} reports normalized confusion matrices (top) and ROC curves (bottom), and we have the following findings:

% To quantify how the pseudo-position signal transfers into separability at the final classifier, we compare four feature settings under the \emph{same} FCNN, split, and preprocessing: 
% (i) \textbf{base\_only}: four hand-crafted scalars (Peak, Rise, Fall, FWHM) only; 
% (ii) \textbf{actual\_only}: KDE-anchored ground-truth positions concatenated with the four scalars; 
% (iii) \textbf{cnn\_pred\_only}: CNN-regressed positions (before calibration) plus the four scalars; 
% (iv) \textbf{calibrated\_only}: PCHIP-calibrated positions (Section~\ref{sec:calib}) plus the four scalars. 
% Fig.~\ref{fig:conf_roc} summarizes row-normalized confusion matrices (top row) and ROC curves (bottom row).

\begin{enumerate}
  \item \textbf{Base features only.} In \texttt{base\_only}, the confusion matrix is highly imbalanced (most samples assigned to the photon class) and the ROC follows the diagonal (\(\mathrm{AUC}\!\approx\!0.514\)), indicating near-chance discrimination. The four scalars alone provide limited class information.
  \item \textbf{KDE Positions.} In \texttt{actual\_only}, both classes concentrate along the confusion-matrix diagonal (dark count \(\sim\!97.6\%\), photon \(\sim\!99.8\%\) correct), and the ROC approaches the upper-left corner (\(\mathrm{AUC}\!\approx\!1.000\)). This aligns with the physical relevance of KDE-anchored positions.
  \item \textbf{Raw CNN positions.} In \texttt{cnn\_pred\_only}, the confusion matrix and ROC (\(\mathrm{AUC}\!\approx\!0.519\)) remain close to \texttt{base\_only}.
  % As shown in Sec.~\ref{sec:pos_regression}, global bias/scale in the raw CNN outputs degrades downstream separability.
  \item \textbf{Calibrated CNN positions.} In \texttt{calibrated\_only}, the confusion matrix is strongly diagonal (dark count \(\sim\!96.8\%\), photon \(\sim\!99.7\%\)) and the ROC is near-perfect (\(\mathrm{AUC}\!\approx\!0.999\)), essentially matching \texttt{actual\_only}. The PCHIP mapping removes global bias/scale while preserving rank, thereby realigning CNN outputs with the KDE-anchored manifold.
\end{enumerate}

These trends, as demonstrated in Fig.~\ref{fig:fcnn_matrix} and associated findings, are consistent with the regression results in Fig.~\ref{fig:cnn_calib}. The CNN outputs being monotonically calibrated make the pseudo position become a stable, physically meaningful discriminator. Specifically, overall classification accuracy and per-class precision/recall/F1 are summarized in Table~\ref{tab:perclass_prf_full}.   Collectively, Fig.~\ref{fig:fcnn_matrix} and Table~\ref{tab:perclass_prf_full} show that (i) scalar waveform statistics are insufficient on their own; (ii) physically anchored positions are highly informative; (iii) raw CNN positions require calibration; and (iv) the proposed PCHIP step restores separability to the level of KDE-anchored positions, supporting the hybrid design. Additionally, the classification accuracy in the scenario of \texttt{calibrated\_only} demonstrates a classification accuracy of 98\%, which shows the effective power of classifying photons and dark counts with the proposed hybrid ML model at \name. 

\begin{table*}[t]
  \centering
  \caption{Per class precision /recall / F1 and overall accuracy on the test set for each feature set.}
  \label{tab:perclass_prf_full}
  \small
  \begin{tabular}{l l
                  S[table-format=1.3]
                  S[table-format=1.3]
                  S[table-format=1.3]
                  c}
    \toprule
    \textbf{Feature set} & \textbf{Class} & {\textbf{Precision}} & {\textbf{Recall}} & {\textbf{F1}} & \textbf{Accuracy} \\
    \midrule
    \multirow{2}{*}{base\_only}
      & dark count   & {0.57} & {0.01} & {0.02} & \multirow{2}{*}{0.57} \\
      & photon & {0.59} & {0.98} & {0.7} &  \\
    \midrule
    \multirow{2}{*}{actual\_only}
      & dark count   & {0.99} & {0.96} & {0.98} & \multirow{2}{*}{0.99} \\
      & photon & {0.98} & {0.99} & {0.99} &  \\
    \midrule
    \multirow{2}{*}{cnn\_pred\_only}
      & dark count   & {0.52} & {0.02} & {0.03} & \multirow{2}{*}{0.58} \\
      & photon & {0.58} & {0.97} & {0.71} &  \\
    \midrule
    \multirow{2}{*}{calibrated\_only}
      & dark count   & {0.99} & {0.97} & {0.98} & \multirow{2}{*}{0.98} \\
      & photon & {0.98} & {0.97} & {0.99} &  \\
    \bottomrule
  \end{tabular}
\end{table*}

% For classifying real photon detection, the laser wavelength is set to 1535 nm.  For dark counts, the laser is turned off, and our SNSPD has a dark count rate of 2 Hz. The collection time for the dark count is 2500 seconds, so that we accumulate a comparable size of data for testing photons and dark counts. 

\subsection{t-SNE Visualization of Discriminative Structure}
To assess whether the different input configurations give rise to separable internal representations, we visualize the \emph{penultimate} FCNN activations by means of t-SNE. 
For each model, let $\phi(x)\in\mathbb{R}^{32}$ denote the 32-dimensional vector immediately before the softmax layer. 
We collect $\phi(x)$ on the test split for four settings\texttt{base\_only} (four scalar features), 
\texttt{actual\_only} (ground truth positions $+$ features), \texttt{cnn\_pred\_only} (raw CNN positions $+$ features), 
and \texttt{calibrated\_only} (PCHIP calibrated positions $+$ features) and embed them into 2D using t-SNE with fixed hyperparameters (perplexity $=30$, learning rate $=200$, iterations $=1000$, fixed seed) to ensure comparability across panels. 
As a compact quantitative summary of class separation, we also report the silhouette coefficient $s\in[-1,1]$~computed on the 2-D embeddings (higher is better).

\paragraph*{Findings}
The \texttt{base\_only} and \texttt{cnn\_pred\_only} embeddings form largely entangled point clouds (silhouette $\approx 0.00$), 
indicating that neither handcrafted scalar features alone nor uncalibrated CNN positions 
induce a discriminative geometry in the network's latent space. 
By contrast, \texttt{actual\_only} yields two compact, well separated clusters with a visible margin (silhouette $\approx 0.35$), confirming that physically accurate positions are strongly informative. 
Importantly, \texttt{calibrated\_only} closely reproduces this structure (silhouette $\approx 0.34$), demonstrating that the monotone calibration step restores the separability that the raw CNN outputs lack. 
These qualitative trends are consistent with the confusion matrices and ROC AUC in Fig.~\ref{fig:fcnn_matrix}, where only \texttt{actual\_only} and \texttt{calibrated\_only} approach near-perfect operating characteristics.

\paragraph*{Remark}
t-SNE is used here as a diagnostic tool; it preserves local neighborhoods but not global distances, and its absolute geometry depends on hyperparameters. 
Nevertheless, under fixed settings the relative ranking across feature sets is stable and supports the efficacy of the calibration stage and classifying ability between photon and dark count with the proposed hybrid ML model in \name.

\begin{figure*}[t]
    \centering
    \includegraphics[width=\linewidth]{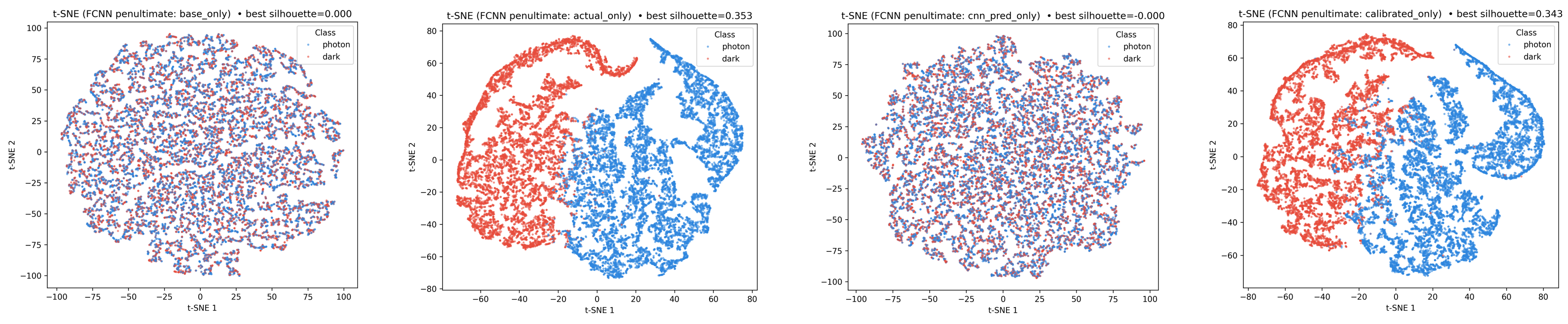}
    \caption{t-SNE of penultimate FCNN features $\phi(x)$ under four input settings. 
  Panels (a)-(d): \texttt{base\_only}, \texttt{actual\_only}, \texttt{cnn\_pred\_only}, and \texttt{calibrated\_only}. 
  Points are colored by class (photon vs.\ dark count). 
  Numbers in titles denote the silhouette coefficient (best over a small hyperparameter grid). 
  Calibrated positions recover the clustered geometry of ground truth positions, whereas scalar features alone or uncalibrated CNN positions remain entangled.}
  \label{fig:tsne_four 
}
    \label{fig:fcnn_tsne}
\vspace{-0.5cm}
\end{figure*}

\section{\name ~Case Study and Performance Evaluation}
\label{sec:casestudy}
This section presents two case studies that utilize the proposed hybrid ML model in \name. First, we utilized \name 
~ model on the elimination of dark count of 20~kM fiber quantum link. Second, we utilize \name 
~ on eliminate the dark count of erbium-based single photon emitter.

\subsection{Case Study 1: 20-km Quantum Link}
\label{sec:20km_snr}
For a practical quantum communication network, optical fiber is primarily used as the medium for photon transmission. Optical fiber is not only the main cause of photon loss but also introduces extra dark count from the environment. The dark count caused by the optical significantly affects the photon stream-level SNR. The SNR in this work is defined as the ratio between real photons and dark counts. This subsection \emph{computes} the SNR improvement delivered by the proposed hybrid ML model on a 20\,km quantum link under controlled illumination. 

\paragraph*{Experimental Setup}
\begin{figure}[h] 
\centering
\includegraphics[width=0.8\linewidth]{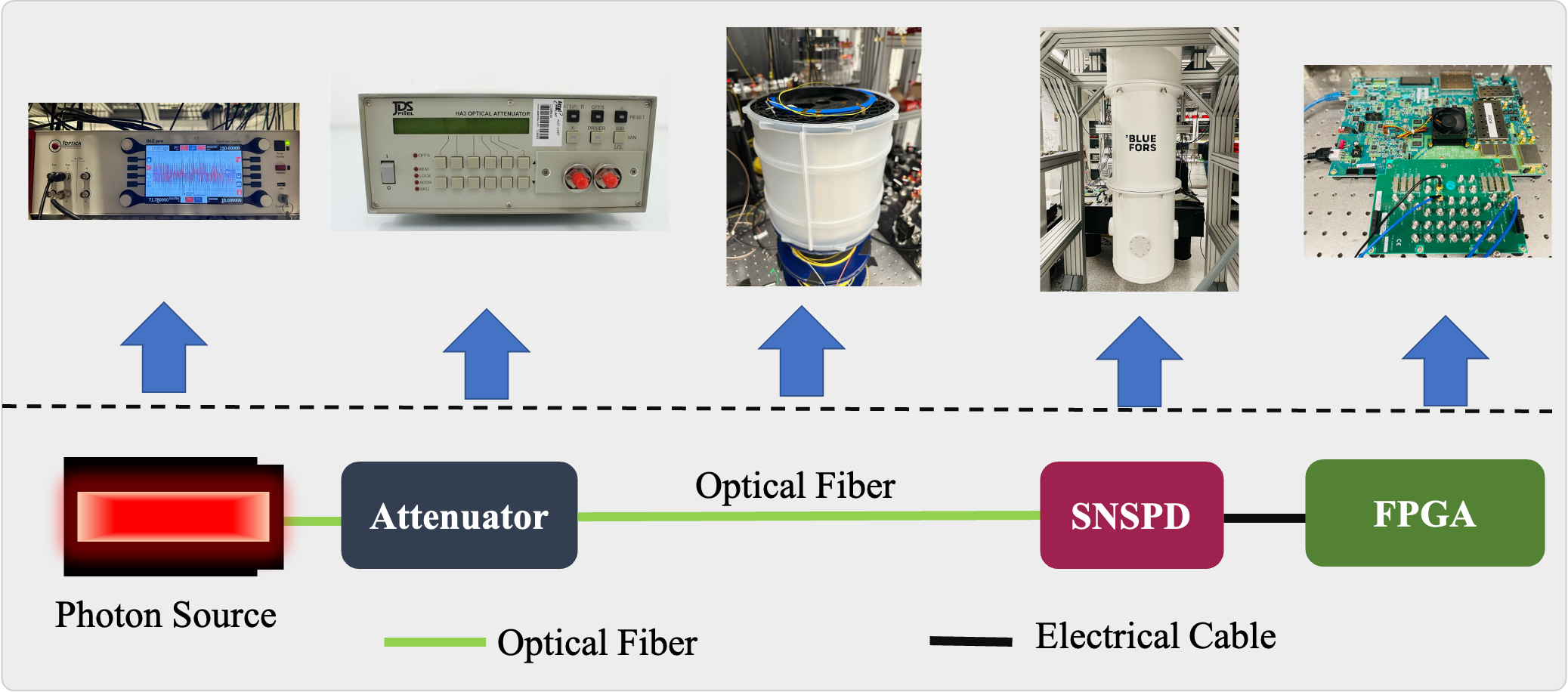}
\caption{Experiment set up for the dark count elimination through 20\,km optical fiber length  }
\label{fig:20kmfiberexp}
\vspace{-0.5cm}
\end{figure}

Fig.\ref{fig:20kmfiberexp} demonstrates the experimental setup for the case study of dark count elimination with 20\,km optical fiber. The experiment setup includes: a photon source as the transmitter, the optical attenuator to maintain the photon rate, a 20 km optical fiber, a photon detection system with SNSDP and FPGA as the receiver. 
At the receiver end of the 20 km span, we acquire SNSPD waveforms and process each event with the fixed pipeline from Sec. IV:
(i) waveform preprocessing; (ii) CNN-based pseudo-position regression; (iii) monotone PCHIP calibration; (iv) FCNN classification (photon vs.\ dark).
In the experiment, we evaluate three ambient regimes by varying background illumination while keeping the photon flux fixed for comparability:
\emph{Dark lab} (low background), \emph{Dim ambient} (moderate background), and \emph{Lights on} (bright room). For reporting, we use steady-state event rates averaged over multiple windows.

\paragraph*{Evaluation Metric}
This section provides the evaluation metric used for the evaluation of the 20km optical fiber experiment,  and the same evaluation metric is applied to the following erbium-based photon emission performance evaluation.  Here, we let $S$ and $B$ denote the photon and dark count rates (events/s). For a given operating point, the classifier attains true-positive rate $\mathrm{TPR}$ and false-positive rate $\mathrm{FPR}$ on the 20\,km dataset, yielding $S'=\mathrm{TPR}\cdot S$, $B'=\mathrm{FPR}\cdot B$ and
% \begin{equation}
% S'=\mathrm{TPR}\cdot S,\qquad B'=\mathrm{FPR}\cdot B,\qquad
% \label{eq:snr_gain_1}
% \end{equation}
% and
\begin{equation}
\mathrm{SNR}'=\frac{S'}{B'}=\underbrace{\Big(\tfrac{\mathrm{TPR}}{\mathrm{FPR}}\Big)}_{G}\cdot\mathrm{SNR}.
\label{eq:snr_gain_2}
\end{equation}
Here, SNR is the ratio between photon and dark count without using the proposed hybrid machine learning model, $S', B', SNR'$, and $G$ are the photon count rate, dark count rate, SNR, and gain factor after applying the proposed hybrid machine learning model ~for the dark count elimination.

\begin{figure}[h] 
\centering
\includegraphics[width=0.7\linewidth]{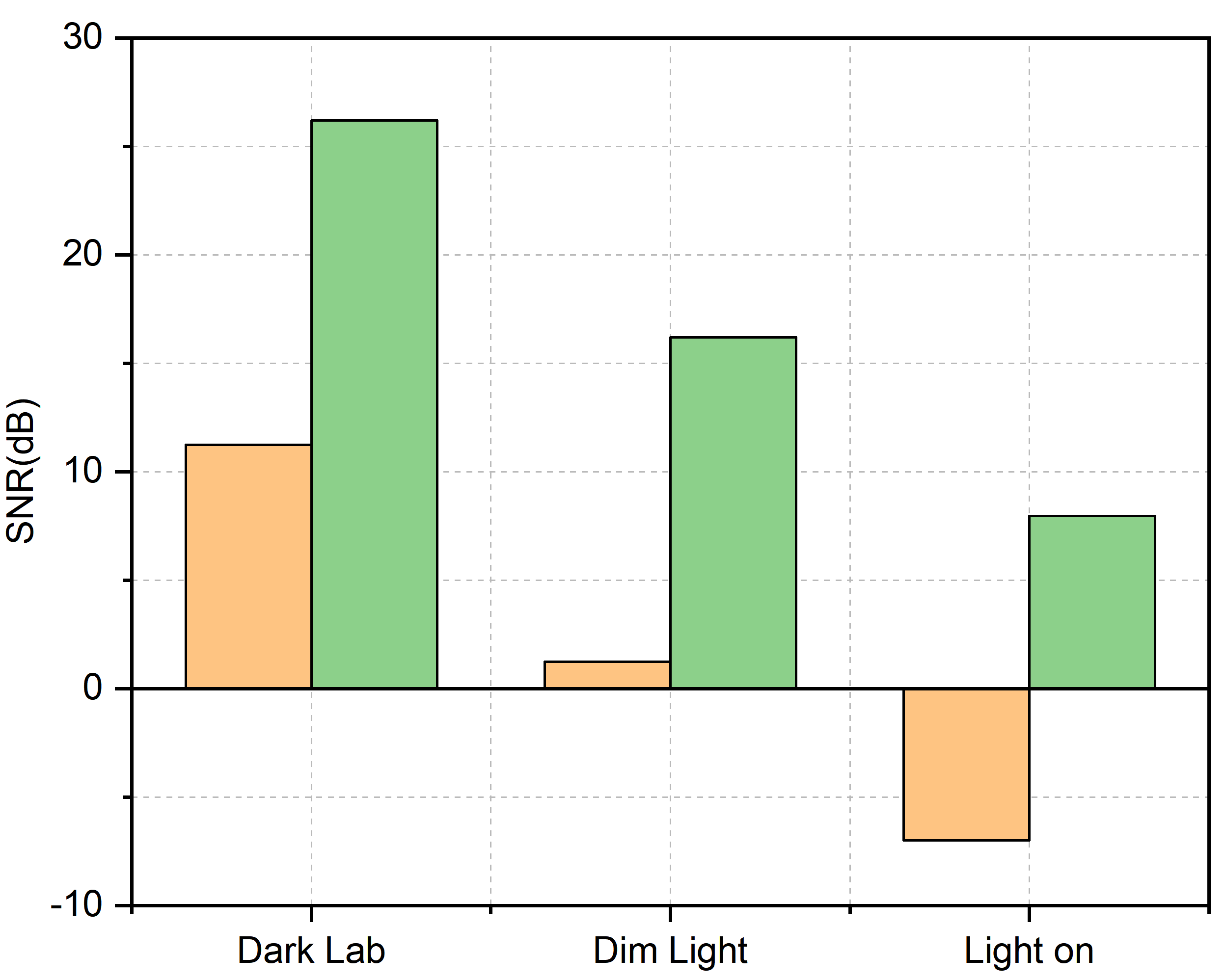}
\caption{20\,km fiber link: stream-level SNR before (orange) and after (green) the proposed hybrid machine learning model under three ambient regimes (Dark Lab, Dim Light, Lights On).}
\label{fig:20kmfiber}
\vspace{-0.8cm}
\end{figure}

\paragraph*{Results analysis}
In the case study, we fix the photon flux at \(S=4000~\mathrm{s^{-1}}\) and sweep the background according to three ambient regimes:
\emph{Dark lab} \(B=300~\mathrm{s^{-1}}\),
\emph{Dim ambient} \(B=3000~\mathrm{s^{-1}}\),
and \emph{Lights on} \(B=20000~\mathrm{s^{-1}}\).
Using the hybrid ML model at a fixed operating point (chosen a priori), we compute the stream-level
\(\mathrm{SNR}\) from the raw photon/dark count event rates and
\(\mathrm{SNR'}\) from the retained/rejected rates after classification, following~\eqref{eq:snr_gain_2}.
Table~\ref{tab:snr_20km} reports the measured \(\mathrm{SNR}\) and \(\mathrm{SNR's}\) in dB for each regime, and
Fig.~\ref{fig:20kmfiber} visualizes the improvement of SNR in dB scales.

As summarized in the table and demonstrated in the figure, we can find that the classifier yields a constant SNR gain of $G\simeq31.2\times (+14.94 dB) $ across all conditions, which is consistent with the calibrated confusion matrices. 
Concretely, \emph{Dark lab} improves from 13.33 (11.25 dB) to 416.00 (26.19 dB); 
\emph{Dim ambient} from 1.33 (1.25 dB) to 41.60 (16.19 dB); and the challenging \emph{Lights on} case from 0.20 (-6.99 dB) to 6.24 (7.95 dB), 
converting a noise-dominated stream into a signal-dominated one. 
These measurements on the 20\,km dataset are consistent with the earlier per-event analysis, 
indicating that the calibrated pseudo-position features translate into robust link-level SNR improvements.

\begin{table*}[t]
  \centering
  \setlength{\tabcolsep}{8pt}
  \renewcommand{\arraystretch}{1.15}
  \caption{20\,km link: stream-level SNR before/after classification. Reported values are the measured numbers (no extra rounding). The nominal gain from the calibrated confusion matrices is \(G\!\approx\!31.2\times\) (\(+14.9\) dB).
  \textit{Note:} The near-constant gain across regimes follows \eqref{eq:snr_gain_2} with
$G=\mathrm{TPR}/\mathrm{FPR}$ at a fixed operating point; TPR/FPR are read from held-out
confusion matrices with a frozen threshold (no post-hoc tuning).}
  \label{tab:snr_20km}
  \begin{tabular}{lcccccc}
    \toprule
    \textbf{Regime} & \(S\) [s\(^{-1}\)] & \(B\) [s\(^{-1}\)] & \(\mathrm{SNR}\) & \(\mathrm{SNR}\) [dB] & \(\mathrm{SNR}'\) & \(\mathrm{SNR}'\) [dB] \\
    \midrule
    Dark lab    & 4000 &   300  & 13.33 &  11.25 & 416.00 & 26.19 \\
    Dim ambient & 4000 &  3000  &  1.33 &   1.25 &  41.60 & 16.19 \\
    Lights on   & 4000 & 20000  &  0.2 &  -6.99 &  6.24 & 7.95 \\
    \bottomrule
  \end{tabular}
\vspace{-0.5cm}
\end{table*}

% \paragraph*{Discussion.}
% (i) The constancy of the gain across regimes is a direct consequence of~\eqref{eq:snr_gain_2}; it indicates that the model acts as a \emph{selective attenuator} on background while preserving almost all photons. 
% (ii) Because $G$ is read off from confusion matrices evaluated on held-out data, the SNR projections here are anchored to measurable classifier operating points rather than tuned thresholds. 
% (iii) Combining this stream-level view with the earlier per-sample metrics provides a complete, reproducible picture of end-to-end performance on a metro-length link.

\noindent
\subsection{Case Study 2: Quantum Emitter} An erbium ion-based quantum emitter is used for the evaluation of \name ~system. In this work, a prototype system is implemented to distinguish and eliminate dark counts from the emitted photon measurement on the quantum emitter. The full details about the experimental setup for the quantum emitter source can be found in ~\cite{gupta2023robust}, and are only briefly described as follows. A single erbium ion in a $Y_{2}O_{3}$ thin-film is coupled to a fiber Fabry-Perot cavity, which is at 10 mK temperature in a dilution refrigerator, as shown in Fig.~\ref{fig:emitter_dilute}. A simplified erbium ion energy level diagram is shown in Fig.~\ref{fig:emitter_dilute}c. The quantum information is encoded in single erbium ions and measured by the emitted photons.
% in the two spin levels in the ground state, and an optical transition at 1535 nm connects one of the lower spin level to an optical excited state. 
When the ion is excited by a laser at 1535 nm, the ion will be excited to the  excited state, and then it decays back to the ground state with emitted photons. Therefore, by detecting the emitted photons, one can infer the quantum information stored in the erbium ion. 
% The role of the fiber FP cavity is to enhance the photoluminescence emission rate of the erbium ion \cite{ahmad2022tunable}, leading to efficient spin-photon entanglement and enabling optical readout of the spin qubit states by counting photons emitted from the ion. 
However, due to the low photon emission of the system, the dark count introduces a significant impact on the readout fidelity of the photon emitter. 
\begin{figure}[h] 
\centering
\includegraphics[width=0.7\linewidth]{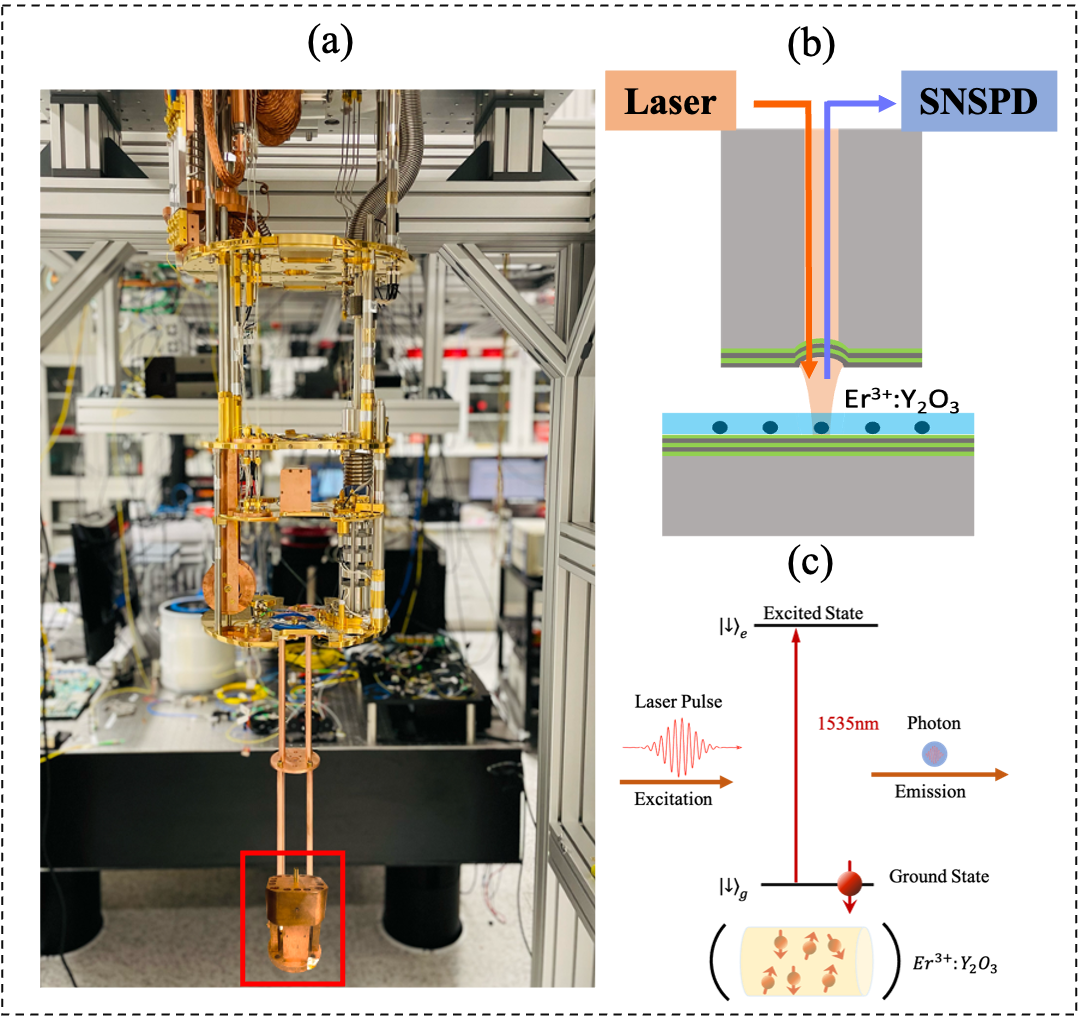}
\caption{Setup for a quantum emitter source based on a single erbium ion. (a) a dilution refrigerator housing the quantum emitter setup; (b) schematics of a fiber Fabry-Perot cavity coupled single erbium ion in $Y_{2}O_{3}$ film; (c) erbium ion energy diagram from the ground state to the excited state.}
\label{fig:emitter_dilute}
\end{figure}
% \vspace{-0.5cm}
Therefore, these experiments apply the proposed hybrid ML model to classify dark counts from the emitted photons and eliminate the effects caused by dark counts. Specifically, this work will demonstrate the ability of \name~ to improve the SNR between emitted photon and dark count.  
\begin{figure}[h] 
\centering
\includegraphics[width=0.7\linewidth]{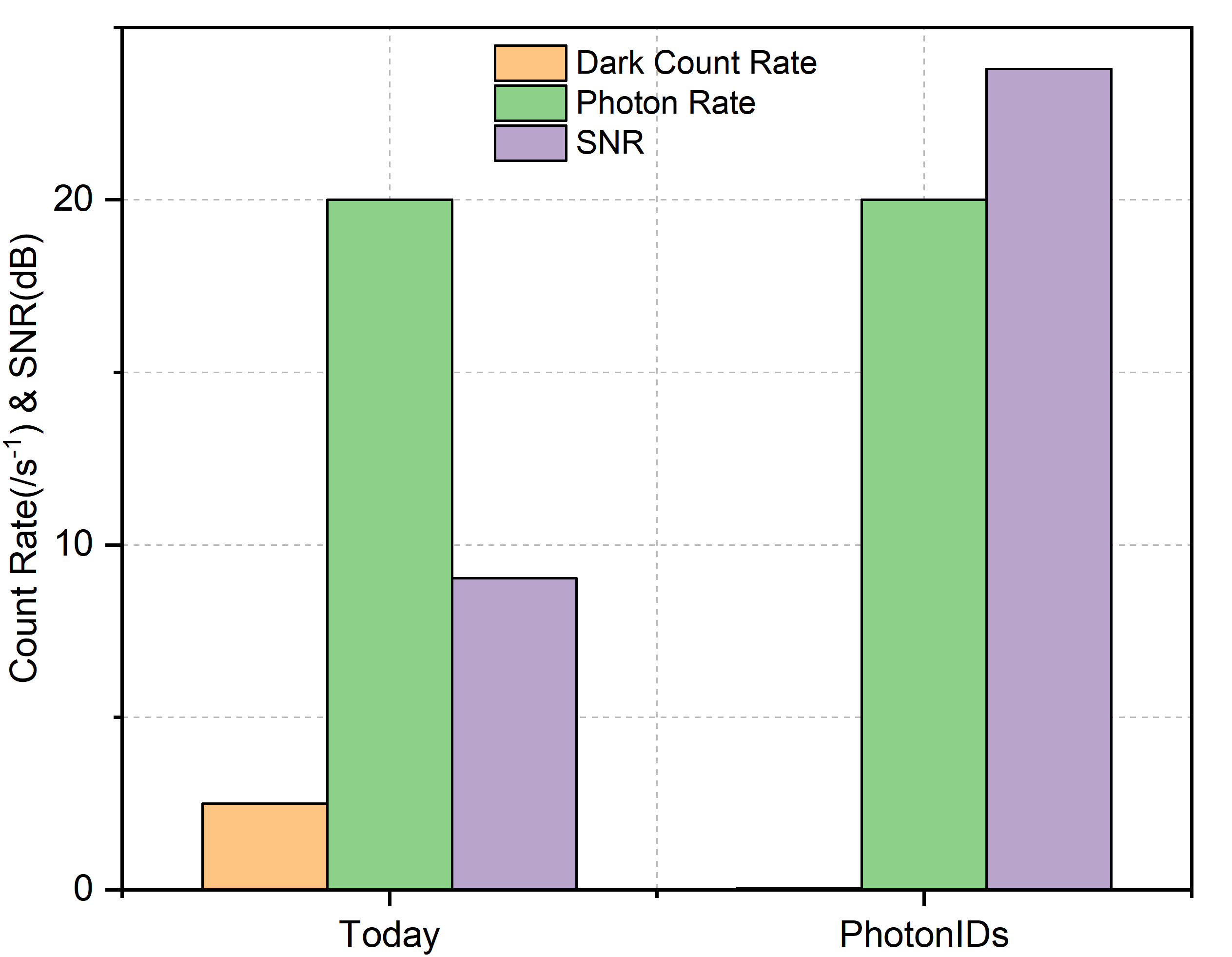}
\caption{Dark-count rate (orange), photon rate (green), and the resulting SNR in dB (purple) for the raw stream (\emph{Today}) and for events accepted by the classifier (\emph{PhotonIDs}). The photon rate is preserved while dark counts are strongly suppressed, improving SNR from $\sim\!9.03$\,dB to $\sim\!23.8$\,dB.}
\label{fig:emitter_dilute}
\vspace{-0.3cm}
\end{figure}
Fig.~\ref{fig:emitter_dilute} summarizes the statistical results before and after applying the hybrid ML model proposed in \name. 
Under identical operating conditions, the raw stream (\emph{Today}) exhibits a photon rate of $\approx\!20~\mathrm{s^{-1}}$ with a small but non-negligible dark-count rate ($\approx\!2.5~\mathrm{s^{-1}}$), yielding an SNR of about $9.03$\,dB. 
After gating by \emph{PhotonIDs}, the retained photon rate remains essentially unchanged (negligible loss), 
while the residual dark counts are suppressed significantly, which translates to an effective SNR of $\approx\!23.8$\,dB.
The $\sim\!14.8$\,dB improvement is consistent with the confusion-matrix--based error rates reported earlier and indicates that the hybrid ML model in \name~ remove the majority of dark counts without sacrificing true photons. This result corroborates the practicality of the \name~ on a laboratory erbium source in addition to the 20\,km link experiments.

\section{Conclusion}
This work proposed \name, the first kind of ML-powered photon identification system with sufficient experimental evaluation. The designed event-driven data acquisition approach enables the real-time event only data extraction and background data suppression.
The proposed hybrid machine learning model yields near-perfect ($R^2\!\approx\!1$) pseudo-position calculation with 1D-CNN regression and PCHIP calibration, and achieves up to 98\% accuracy in classifying the photon and dark count with the proposed compact FCNN classifier. Importantly, the practical case study of \name ~on the scenarios of 20\,km quantum link and erbium-based photon emitter demonstrates efficient performance of dark count elimination with 31.2 times(14.9dB) improvement on SNR. The proposed \name~ provides a practical path toward dark count-free quantum communication at scale. 

\section{Acknowledgements}
\label{sec:ACK}
% \vspace*{-5pt}
This work is supported by PHOTONIDS, Inc., and the U.S. DOE Office of Science-Basic Energy Sciences, under Contract No. DEAC02-06CH11357.

\end{document}